\documentclass[a4paper,11pt]{article}
\pdfoutput=1
\usepackage{jheppub}

\usepackage{amssymb,amsmath,amsfonts}
\usepackage[normalem]{ulem}
\usepackage[utf8x]{inputenc}
\usepackage{slashed}
\usepackage{graphicx}
\usepackage{tabularx}
\usepackage{here}
\usepackage{color}
\usepackage{csquotes} 
\usepackage{comment}
\usepackage{mathrsfs}
\usepackage{float}
\usepackage{ascmac}
\usepackage{multirow}
\usepackage{longtable}
\usepackage{bm}
\usepackage{ulem}

\usepackage[italicdiff]{physics}
\usepackage{caption}
\newcounter{c1}
\newcounter{c}
\newcounter{c2}
\newcounter{c3}
\newcounter{c4}
\makeatletter
\newcommand*\rel@kern[1]{\kern#1\dimexpr\macc@kerna}
\newcommand*\widebar[1]{%
  \begingroup
  \def\mathaccent##1##2{%
    \rel@kern{0.8}%
    \overline{\rel@kern{-0.8}\macc@nucleus\rel@kern{0.2}}%
    \rel@kern{-0.2}%
  }%
  \macc@depth\@ne
  \let\math@bgroup\@empty \let\math@egroup\macc@set@skewchar
  \mathsurround\z@ \frozen@everymath{\mathgroup\macc@group\relax}%
  \macc@set@skewchar\relax
  \let\mathaccentV\macc@nested@a
  \macc@nested@a\relax111{#1}%
  \endgroup
}
\makeatother

\numberwithin{equation}{section}

\preprint{
\begin{minipage}{5cm}
\small
\flushright
EPHOU-26-01\\
KYUSHU-HET-352
\end{minipage}}

\title{
Yukawa Textures with Enhanced Symmetries\\ in Heterotic Calabi-Yau Compactifications
}

\author{Jun Dong$^{1}$,} 
\author{Tatsuo Kobayashi$^{1}$,} 
\author{Shuhei Miyamoto$^{1}$, and} 
\author{Hajime Otsuka$^{2,3}$}
\affiliation{
$^1$Department of Physics, Hokkaido University, Sapporo 060-0810, Japan\\
$^2$Department of Physics, Kyushu University, 744 Motooka, Nishi-ku, Fukuoka 819-0395, Japan\\
$^3$Quantum and Spacetime Research Institute (QuaSR), Kyushu University, 744 Motooka, Nishi-ku, Fukuoka, 819-0395, Japan
}
\emailAdd{j-dong@particle.sci.hokudai.ac.jp}
\emailAdd{kobayashi@particle.sci.hokudai.ac.jp}
\emailAdd{s-miyamoto@particle.sci.hokudai.ac.jp}
\emailAdd{otsuka.hajime@phys.kyushu-u.ac.jp}

\abstract{
We clarify the structure of Yukawa couplings and mass matrices for matter fields in heterotic string theory on smooth Calabi-Yau threefolds with standard embedding. 
The topological structure of Calabi-Yau threefolds leads to interesting Yukawa textures that cannot be derived from group-theoretical symmetries, e.g., the so-called Weinberg texture in the case of two generations of matter fields. 
Furthermore, we find that a $U(2)$ flavor symmetry, which plays an important role in controlling higher-dimensional operators in the Standard Model effective field theory, emerges at specific loci in the moduli space of multi-Higgs fields. Small perturbations around these loci generate semi-realistic patterns of quark masses and mixings. 
}

\makeatletter
\gdef\@fpheader{}
\makeatother

\begin{document}

\maketitle

\section{Introduction}
\label{sec:Intro}

String theory is a promising candidate for a unified theory of all fundamental interactions including gravity and matter fields such as quarks, leptons, and the Higgs boson.
It is expected that string theory can solve various mysteries in particle physics and cosmology.
Superstring theory predicts the existence of a six-dimensional compact space in addition to our four-dimensional spacetime. 
All properties of particle physics and cosmology in four-dimensional low-energy effective field theory are determined by the details of string compactifications.

The presence or absence of specific particle modes is governed by the behavior of the corresponding string states in the compact space. 
In addition, their allowed couplings as well as forbidden ones are controlled by string compactifications.
That is known as {\it stringy coupling selection rules}. 
Moreover, the magnitudes of the couplings depend on geometrical parameters of the compact space. 
Stringy coupling selection rules can often be understood in terms of group theory.
For example, heterotic orbifold models and intersecting/magnetized D-brane models lead to discrete flavor symmetry groups \cite{Dijkgraaf:1987vp,Kobayashi:2004ya,Kobayashi:2006wq,Beye:2014nxa,Abe:2009vi,Berasaluce-Gonzalez:2012abm,Marchesano:2013ega}.
These discrete flavor symmetries are useful to understand the origin of flavor structures such as mass hierarchies of quarks and leptons and their mixing angles. 
(See Refs.~\cite{Altarelli:2010gt,Ishimori:2010au,Kobayashi:2022moq,Hernandez:2012ra,King:2013eh,King:2014nza,Petcov:2017ggy} for bottom-up approaches based on discrete flavor symmetries.)

Also, in certain classes of heterotic orbifold models and intersecting/magnetized D-brane models, modular symmetries behave as flavor symmetries \cite{Lauer:1989ax,Lauer:1990tm,Lerche:1989cs,Ferrara:1989qb,Baur:2019kwi,Nilles:2020nnc,Kobayashi:2018rad,Kobayashi:2018bff,Kikuchi:2021ogn,Ohki:2020bpo,Kikuchi:2020frp,Kikuchi:2020nxn,Almumin:2021fbk,Ishiguro:2023jqb,Kikuchi:2023awe,Kobayashi:2017dyu}. 
In addition, the modular symmetries act as flavor symmetries in heterotic string theory on Calabi-Yau (CY) manifolds \cite{Strominger:1990pd,Candelas:1990pi,Ishiguro:2020nuf,Ishiguro:2021ccl}. 
These modular flavor symmetries also provide a useful framework for understanding the origin of flavor structure.
(See Refs.~\cite{Feruglio:2017spp,Kobayashi:2018vbk,Penedo:2018nmg,Novichkov:2018nkm,Kobayashi:2018scp} for bottom-up approaches based on modular flavor symmetries.\footnote{See for reviews of the bottom-up approaches to modular flavor models \cite{Kobayashi:2023zzc,Ding:2023htn}.})

On the other hand, some of the stringy selection rules cannot be understood in terms of group-theoretical symmetries. 
For example, such stringy selection rules in heterotic orbifold models were studied in Refs.~\cite{Font:1988nc,Cvetic:1987qx,Kobayashi:2011cw}. 
In particular, conjugacy classes of space groups play an essential role in determining the selection rules in heterotic orbifold models \cite{Dixon:1986jc,Hamidi:1986vh,Dixon:1986qv,Kobayashi:1990mc,Kobayashi:1991rp,Kobayashi:1995py,Heckman:2024obe,Kaidi:2024wio,Kobayashi:2025ocp}. 
Similar structures arise in magnetized D-brane models on toroidal orbifolds \cite{Kobayashi:2024yqq,Funakoshi:2024uvy}. 
These structures correspond to hypergroups mathematically, and their selection rules are non-invertible. 
Recently, these selection rules, e.g. $\mathbf{Z}_2$ gauging of $\mathbf{Z}_N$ symmetries, have been applied to particle physics, leading to interesting results, e.g. novel Yukawa textures of quarks and leptons \cite{Kobayashi:2024cvp,Kobayashi:2025znw,Kobayashi:2025ldi,Jiang:2025psz}, axion-less solutions to the strong CP problem~\cite{Liang:2025dkm,Kobayashi:2025thd,Kobayashi:2025rpx}, radiative seesaw models~\cite{Kobayashi:2025cwx,Nomura:2025sod,Chen:2025awz,Okada:2025kfm}, suppression of flavor changing neutral currents~\cite{Nakai:2025thw}, and the stability of dark matter~\cite{Suzuki:2025oov,Kobayashi:2025lar}, which cannot be derived from group-theoretical symmetries.

Furthermore, heterotic string theory on CY manifolds leads to non-trivial selection rules. 
These selection rules are not governed by group-theoretical considerations, but rather by topological properties of the CY threefolds. 
When we consider the so-called standard embedding~\cite{Candelas:1985en}, 
one can extract the effective action of matter fields from the corresponding moduli fields.\footnote{For the scenario of non-standard embedding, the quark masses and mixings have been studied in e.g., Refs.~\cite{Constantin:2024yxh,Constantin:2025vyt} using machine-learning techniques. The hierarchical structure of quark masses has been obtained by using Froggatt–Nielsen mechanism~\cite{Froggatt:1978nt}.}
Then, the allowed couplings of matter fields are determined by intersection numbers of the corresponding cycles. 
Recently, such selection rules have been systematically classified for complete intersection Calabi-Yau (CICY) manifolds with five or fewer moduli \cite{Dong:2025pah}. 
The structure of Yukawa textures for matter fields was also studied. 
It is important to further investigate the phenomenological implications of these textures.
That is our purpose. 
In this paper, we study fermion mass hierarchies arising from heterotic Calabi-Yau compactifications with standard embedding with particular emphasis on the structure of Yukawa textures.

This paper is organized as follows. 
In section~\ref{sec:CY}, we review the low-energy effective action of matter fields, in particular, their K\"ahler potential and Yukawa couplings, in heterotic string theory on CY threefolds with standard embedding. 
After reviewing the coupling selection rules of matter fields in section \ref{sec:selection-rule}, we explicitly analyze their Yukawa textures and mass hierarchies for CICYs with $h^{1,1}=2$ in section~\ref{sec:2-moduli} and $h^{1,1}=3$ in section~\ref{sec:3-moduli}. 
Realistic patterns of quark masses and mixings are demonstrated in section~\ref{sec:pheno}. 
Section~\ref{sec:con} is devoted to the conclusions. 
In Appendix~\ref{app:h11_3}, we summarize the Yukawa textures of matter fields for CICYs with $h^{1,1}=3$.

\section{Calabi-Yau compactifications and low energy effective field theory}
\label{sec:CY}

In this section, we briefly review the low-energy effective action of heterotic string theory on smooth CY threefolds. 
In particular, we focus on the standard embedding, in which the four-dimensional gauge theory is described by $E_6\times E_8'$. 
Since the $SU(3)\subset E_8$ bundle is identified with the tangent bundle of CY threefolds, the $\mathbf{27}$ fundamental and $\overline {\mathbf{27}}$ anti-fundamental representations of $E_6$ are in one-to-one correspondence with K\"ahler and complex structure moduli of CY threefolds, respectively. 
Hence, the low-energy effective action of these matter fields can be derived directly from those of corresponding moduli fields. 

Following Ref.~\cite{Dixon:1989fj}, let us consider the effective action of the matter fields $\mathbf{27}^a$, i.e., fundamental representation of $E_6$, represented by $A^a$. 
When their field values are small, i.e., $|A^a| \ll 1$, the matter K\"ahler metric is described by
\begin{align}
\label{eq:Kahler-metric}
   K^{(\mathbf{27})}_{a\bar{b}} &= e^{\frac{1}{3}(K_{\rm cs}-K_{\rm ks})}(K_{\rm ks})_{a\bar{b}},
\end{align}
where $K_{\rm ks}$ and $K_{\rm cs}$ respectively denote the K\"ahler potential of K\"ahler and complex structure moduli:
\begin{align}
\label{eq:Kahler-potential}
    K_{\rm ks} &= - \ln \biggl[\frac{1}{6}\sum_{a,b,c}\kappa_{abc} (T^a+\bar{T}^a)(T^b+\bar{T}^b)(T^c+\bar{T}^c)\biggl],
        \nonumber\\
    K_{\rm cs} &= - \ln \biggl[\frac{i}{6}\sum_{i,j,k}\kappa_{ijk}(U^i-\bar{U}^i)(U^j-\bar{U}^j)(U^k-\bar{U}^k)\biggl],
\end{align}
with $\kappa_{abc}$ and $\kappa_{ijk}$ corresponding to intersection numbers. 
Here, we assume the large K\"ahler structure and complex structure regime such that loop and quantum corrections are sufficiently suppressed compared to the classical terms considered here. 
However, taking the K\"ahler moduli to be arbitrarily large is not consistent with the observed value of four-dimensional gauge coupling. 
From the low-energy effective action of Yang-Mills fields:
\begin{align}
S_{\rm bos}&\supset-\frac{1}{2g_{10}^2} \int_{M^{(10)}} e^{-2\phi_{10}} 
{\rm tr}(F\wedge \ast F),
\label{eq:heterob}
\end{align}
with $g_{10}^2=2(2\pi)^7(\alpha')^3$ being the ten-dimensional gauge coupling, 
the four-dimensional gauge coupling $g_4$ is written in terms of the string coupling $g_s = e^{\langle \phi_{10}\rangle}$ and the CY volume in units of string length $l_s=2\pi \sqrt{\alpha}$, i.e., ${\rm Vol}({\cal M})={\cal V}\,l_s^6$:
     \begin{align}
     \label{eq:g-V}
         \alpha^{-1} = \frac{4\pi}{g_4^2} = e^{-2\langle \phi_{10}\rangle}\frac{4\pi {\rm Vol}({\cal M})}{g_{10}^2} = g_s^{-2} {\cal V},
     \end{align}
     where 
     \begin{align}
         {\cal V}=\frac{1}{48}\sum_{a,b,c}\kappa_{abc}(T^a+\bar{T}^a)(T^b+\bar{T}^b)(T^c+\bar{T}^c).
     \end{align}
Hence, the four-dimensional gauge coupling at the scale of grand unified theory (GUT), $\alpha^{-1}(M_{\rm GUT}) \simeq 25$ with $M_{\rm GUT} \simeq 2\times 10^{16}$\,GeV, puts an upper bound on the CY volume:
     \begin{align}
         {\cal V}=g_s^2 \alpha^{-1} \lesssim 25,
     \end{align}
assuming a perturbative string coupling $g_s\lesssim 1$.

Furthermore, the three-point couplings of $\mathbf{27}$, i.e., the Yukawa interactions, are described by the following superpotential in terms of the intersection numbers~\cite{Strominger:1985it,Candelas:1987se}:
\begin{align}
     W &= \frac{1}{6} \kappa_{abc}A^aA^bA^c.
\end{align}
Then, physical Yukawa couplings can be obtained after canonically normalizing the matter fields. 
Note that the fields $A^a$ include quarks and leptons as well as Higgs fields. 
As discussed above, the existence of hierarchical Yukawa structures depends on the texture of the holomorphic Yukawa couplings. 
For concreteness, we restrict our analysis to CICY classified in Refs.~\cite{Candelas:1987kf,Candelas:1987du,Green:1987cr,He:1990pg,Gagnon:1994ek}. 
This class of CY threefolds is embedded in an ambient space, i.e., a product of complex projective spaces $\mathbb{P}^{n_1}\times\cdots\mathbb{P}^{n_m}$. The following configuration matrix characterizes the CICYs:
\begin{align}
\begin{matrix}
\mathbb{P}^{n_1}\\
\mathbb{P}^{n_2}\\
\vdots\\
\mathbb{P}^{n_m}\\
\end{matrix}
\begin{bmatrix}
q_1^1 & q_2^1 & \cdots & q_R^1\\
q_1^2 & q_2^2 & \cdots & q_R^2\\
\vdots & \vdots & \ddots & \vdots\\
q_1^m & q_2^m & \cdots & q_R^m\\
\end{bmatrix},
\label{eq:conf_matrix}
\end{align}
which means that $n^l$ and the positive integers $q_r^l$ ($r=1,...,R$, $l=1,...,m$), specifying $R$ multi-degrees of homogeneous polynomials, are restricted to satisfy the so-called CY condition:
\begin{align}
    \sum_{r=1}^R q_r^l &= n_l +1\quad (\forall l),
    \qquad
    \sum_{l=1}^m n_l = R + 3.
\end{align}
Among the total of 7890 CICYs\footnote{Note that some CICYs share the same topological properties \cite{Anderson:2008uw}.}, in this paper, we focus on the CICYs whose second cohomology descends from that of the ambient space, namely the so-called favorable CICYs.

\section{Yukawa textures and mass hierarchies}
\label{sec:CICY}

Here, we study Yukawa textures and mass hierarchies arising from heterotic string theory on CICYs with $h^{1,1} =2,3,4$.

\subsection{Coupling selection rules and Yukawa textures}
\label{sec:selection-rule}

In this section, we review the results of Ref.~\cite{Dong:2025pah} concerning coupling selection rules and Yukawa textures derived from heterotic string theory on CICYs with $h^{1,1}=2,3,4$. 
In Ref.~\cite{Dong:2025pah}, CICYs were classified by their intersection numbers $\kappa_{abc}$. 
In particular, it was systematically classified which intersection numbers $\kappa_{abc}$ vanish. 
Table \ref{tab:h11=2} shows this classification for $h^{1,1}=2$. 
Note that Type 4 was missed in Ref.~\cite{Dong:2025pah}.
Table \ref{tab:type-h11=2} lists concrete examples of CICYs, identified by their numbering in Ref.~\cite{CICY}.
In addition, Table \ref{tab:h11=2-texture} shows possible Yukawa textures for two generations $A^{1,2}$ under the assumption that one of $A^1$ and $A^2$ corresponds to the Higgs field. 
Note that the texture of Type 2 with the Higgs field $A^1$ and the texture for Type 3 with the Higgs field $A^2$ cannot be realized by group-theoretical symmetries. 
For example, it is impossible to assign consistent $U(1)$ charges to the two generations to reproduce this texture. 
It is known as the Weinberg texture in the down sector \cite{Weinberg:1977hb}, which can explain the Cabibbo angle by the mass ratio between $m_d$ and $m_s$ as $\sqrt{m_d/m_s}$.

\begin{table}[H]
    \centering
       \caption{Types of prepotential for CICYs with $h^{1,1}=2$. Vanishing $\kappa_{abc}=0$ are shown.}
       \label{tab:h11=2}
    \begin{tabular}{|c||c|}
    \hline
    Type & $\kappa_{abc}=0$ \\ \hline \hline 
    1 & none \\
      \hline
      2 & $\kappa_{111}=0$ \\
      \hline
      3 & $\kappa_{111}=\kappa_{112}=0$ \\
      \hline
      4 & $\kappa_{111}=\kappa_{222}=0$ \\
      \hline
    \end{tabular}
  \end{table}

\begin{longtable}[h]{|c||p{12cm}|}
     \caption{Types of concrete CICYs with $h^{1,1}=2$.}  \label{tab:type-h11=2}\\
        \hline
        Type & \multicolumn{1}{c|}{Number of CICY}\\
        \hline
        \addtocounter{c3}{1}
        \arabic{c3} &
        $ 7644,$ 
        $ 7726 ,$ 
        $ 7759 ,$ 
        $ 7761 ,$ 
        $ 7799 ,$ 
        $ 7809 ,$ 
        $ 7863 $ 
        \\
        \hline
        \addtocounter{c3}{1}
        \arabic{c3} &
        $7643,$
        $ 7668,$
        $ 7725,$
        $ 7758,$
        $ 7807,$
        $ 7808,$
        $ 7821,$
        $ 7833,$
        $ 7844,$
        $ 7853,$
        $ 7868,$
        $ 7883$
        \\
        \hline
        \addtocounter{c3}{1}
        \arabic{c3} &
        $7806,$
        $ 7816,$
        $ 7817,$
        $ 7819,$
        $ 7822,$
        $ 7823,$
        $ 7840,$
        $ 7858,$
        $ 7867,$
        $ 7869,$
        $ 7873,$
        $ 7882,$
        $ 7885,$
        $7886,$
        $ 7887,$
        $ 7888 $
        \\
        \hline
        \addtocounter{c3}{1}
        \arabic{c3} &
        $7884$
        \\
        \hline
    \end{longtable}

    \begin{table}[H]
        \centering
        \caption{Yukawa textures from CICYs with $h^{1,1}=2$.}
        \label{tab:h11=2-texture}
          \begin{tabular}{|c|c|c|}
            \hline
            Type \textbackslash 
            Higgs field
            & $A^1$ & $A^2$\\
            \hline \hline
            \addtocounter{c4}{1}
            Type \arabic{c4} 
            & $\pmqty{
                 *  & 
                 * \\ 
                 *  & 
                 * \\ 
            }$
            & $\pmqty{
                 *  & 
                 * \\ 
                 *  & 
                 * \\ 
            }$
            \\
            \hline
            \addtocounter{c4}{1}
            Type \arabic{c4} 
            & $\pmqty{
                 0  & 
                 * \\ 
                 *  & 
                 * \\ 
            }$
            & $\pmqty{
                 *  & 
                 * \\ 
                 *  & 
                 * \\ 
            }$
            \\
            \hline
            \addtocounter{c4}{1}
            Type \arabic{c4} 
            & $\pmqty{
                 0  & 
                 0 \\ 
                 0  & 
                 * \\ 
            }$
            & $\pmqty{
                 0  & 
                 * \\ 
                 *  & 
                 * \\ 
            }$
            \\
            \hline
            \addtocounter{c4}{1}
            Type \arabic{c4} 
            & $\pmqty{
                 0  & 
                 * \\ 
                 *  & 
                 * \\ 
            }$
            & $\pmqty{
                 *  & 
                 * \\ 
                 *  & 
                 0 \\ 
            }$
            \\
            \hline
        \end{tabular}
     \end{table}

Similarly, we can classify CICYs with $h^{1,1}=3$ according to their intersection numbers. 
The results are summarized in Tables \ref{tab:h11=3},~\ref{tab:type-h11=3}, and \ref{tab:h11=3-texture}, which are organized in the same manner as the corresponding tables for $h^{1,1}=2$. 
From this classification, we obtain non-trivial Yukawa textures. It is notable that some of these textures cannot be reproduced by conventional group-theoretical symmetries.

\begin{table}[H]
    \centering
    \caption{Types of prepotential for CICYs with $h^{1,1}=3$. Vanishing $\kappa_{abc}$ are shown.}
    \label{tab:h11=3}
    \begin{tabular}{|c||c|}
    \hline
    Type & $\kappa_{abc}=0$ \\ \hline \hline
      1 & $\kappa_{111}, \kappa_{222}, \kappa_{333}$ \\
      \hline
       2 & $\kappa_{111}, \kappa_{222}$ \\
      \hline
      3 & $\kappa_{111}, \kappa_{112}, \kappa_{113}$ \\
      \hline
       4 & $\kappa_{111}, \kappa_{112}, \kappa_{113}, \kappa_{222}$ \\
      \hline
       5 & $\kappa_{111}$ \\
      \hline
       6 & none \\
      \hline
       7 & $\kappa_{111}, \kappa_{112}, \kappa_{113} ,\kappa_{122} ,\kappa_{222} ,\kappa_{223}$ \\
      \hline
       8 & $\kappa_{111}, \kappa_{112}, \kappa_{113} ,\kappa_{122} ,\kappa_{222}$ \\
      \hline
       9 & $\kappa_{111}, \kappa_{112}, \kappa_{113}, \kappa_{222}, \kappa_{333}$ \\
      \hline
       10 & $\kappa_{111}, \kappa_{112}, \kappa_{113}, \kappa_{122} ,\kappa_{222}, \kappa_{333}$ \\
      \hline
       11 & $\kappa_{111}, \kappa_{112}, \kappa_{113}, \kappa_{122}, \kappa_{222}, \kappa_{223}, \kappa_{333}$ \\
      \hline
    \end{tabular}
  \end{table}

\begin{longtable}[h]{|c||p{12cm}|}
     \caption{Types of concrete CICYs with $h^{1,1}=3$.}  \label{tab:type-h11=3}\\
        \hline
        Type & \multicolumn{1}{c|}{Number of CICY}\\
        \hline
        \addtocounter{c2}{1}
        \arabic{c2} &
        $ 5299 ,$ 
        $ 6971 ,$ 
        $ 7580 ,$ 
        $ 7581 ,$ 
        $ 7669 ,$ 
        $ 7729 ,$ 
        $ 7846 $ 
        \\
        \hline
        \addtocounter{c2}{1}
        \arabic{c2} &
        $ 6220 ,$ 
        $ 6555 ,$ 
        $ 6827 ,$ 
        $ 6972 ,$ 
        $ 7143 ,$ 
        $ 7235 ,$ 
        $ 7240 ,$ 
        $ 7365 ,$ 
        $ 7366 ,$ 
        $ 7369 ,$ 
        $ 7486 ,$ 
        $ 7534 ,$ 
        $ 7583 ,$ 
        $ 7612 ,$ 
        $ 7646 ,$ 
        $ 7698 ,$ 
        $ 7762 ,$ 
        $ 7791 $ 
        \\
        \hline
        \addtocounter{c2}{1}
        \arabic{c2} &
        $ 6771 ,$ 
        $ 7036 ,$ 
        $ 7208 ,$ 
        $ 7530 ,$ 
        $ 7563 ,$ 
        $ 7566 ,$ 
        $ 7571 ,$ 
        $ 7578 ,$ 
        $ 7588 ,$ 
        $ 7626 ,$ 
        $ 7631 ,$ 
        $ 7635 ,$ 
        $ 7636 ,$ 
        $ 7638 ,$ 
        $ 7647 ,$ 
        $ 7648 ,$ 
        $ 7679 ,$ 
        $ 7717 ,$ 
        $ 7721 ,$ 
        $ 7734 ,$ 
        $ 7747 ,$ 
        $ 7781 ,$ 
        $ 7842 $ 
        \\
        \hline
        \addtocounter{c2}{1}
        \arabic{c2} &
        $ 7069 ,$ 
        $ 7316 ,$ 
        $ 7317 ,$ 
        $ 7452 ,$ 
        $ 7464 ,$ 
        $ 7485 ,$ 
        $ 7556 ,$ 
        $ 7558 ,$ 
        $ 7561 ,$ 
        $ 7562 ,$ 
        $ 7570 ,$ 
        $ 7584 ,$ 
        $ 7585 ,$ 
        $ 7587 ,$ 
        $ 7610 ,$ 
        $ 7627 ,$ 
        $ 7645 ,$ 
        $ 7676 ,$ 
        $ 7697 ,$ 
        $ 7710 ,$ 
        $ 7711 ,$ 
        $ 7720 ,$ 
        $ 7730 ,$ 
        $ 7752 ,$ 
        $ 7755 ,$ 
        $ 7763 ,$ 
        $ 7798 ,$ 
        $ 7802 ,$ 
        $ 7824 ,$ 
        $ 7843 ,$ 
        $ 7847 ,$ 
        $ 7854 $ 
        \\
        \hline
        \addtocounter{c2}{1}
        \arabic{c2} &
        $ 7071 ,$ 
        $ 7144 ,$ 
        $ 7237 ,$ 
        $ 7370 ,$ 
        $ 7488 ,$ 
        $ 7586 $ 
        \\
        \hline
        \addtocounter{c2}{1}
        \arabic{c2} &
        $ 7242 $ 
        \\
        \hline
        \addtocounter{c2}{1}
        \arabic{c2} &
        $ 7450 ,$ 
        $ 7481 ,$ 
        $ 7484 ,$ 
        $ 7555 ,$ 
        $ 7560 ,$ 
        $ 7579 ,$ 
        $ 7661 ,$ 
        $ 7662 ,$ 
        $ 7677 ,$ 
        $ 7694 ,$ 
        $ 7707 ,$ 
        $ 7714 ,$ 
        $ 7735 ,$ 
        $ 7745 ,$ 
        $ 7746 ,$ 
        $ 7753 ,$ 
        $ 7760 ,$ 
        $ 7769 ,$ 
        $ 7776 ,$ 
        $ 7779 ,$ 
        $ 7780 ,$ 
        $ 7788 ,$ 
        $ 7789 ,$ 
        $ 7792 ,$ 
        $ 7795 ,$ 
        $ 7797 ,$ 
        $ 7812 ,$ 
        $ 7834 ,$ 
        $ 7836 ,$ 
        $ 7841 ,$ 
        $ 7845 ,$ 
        $ 7848 ,$ 
        $ 7851 ,$ 
        $ 7865 ,$ 
        $ 7871 ,$ 
        $ 7872 ,$ 
        $ 7874 ,$ 
        $ 7877 ,$ 
        $ 7881 $ 
        \\
        \hline
        \addtocounter{c2}{1}
        \arabic{c2} &
        $ 7465 ,$ 
        $ 7466 ,$ 
        $ 7565 ,$ 
        $ 7576 ,$ 
        $ 7577 ,$ 
        $ 7637 ,$ 
        $ 7678 ,$ 
        $ 7680 ,$ 
        $ 7712 ,$ 
        $ 7713 ,$ 
        $ 7756 ,$ 
        $ 7774 ,$ 
        $ 7782 ,$ 
        $ 7783 ,$ 
        $ 7787 ,$ 
        $ 7801 ,$ 
        $ 7804 ,$ 
        $ 7832 ,$ 
        $ 7838 ,$ 
        $ 7855 ,$ 
        $ 7866 ,$ 
        $ 7876 $ 
        \\
        \hline
        \addtocounter{c2}{1}
        \arabic{c2} &
        $ 7708 ,$ 
        $ 7727 ,$ 
        $ 7728 ,$ 
        $ 7831 ,$ 
        $ 7870 $ 
        \\
        \hline
        \addtocounter{c2}{1}
        \arabic{c2} &
        $ 7875 $ 
        \\
        \hline
        \addtocounter{c2}{1}
        \arabic{c2} &
        $ 7880 $ 
        \\
        \hline
    \end{longtable}

    \begin{table}[H]
        \centering
        \caption{Yukawa textures from CICYs with $h^{1,1}=3$.}
        \label{tab:h11=3-texture}
          \begin{tabular}{|c|c|c|c|}
            \hline
            Type \textbackslash Higgs field & $A^1$ & $A^2$ & $A^3$\\
            \hline \hline
            \addtocounter{c}{1}
            Type \arabic{c} 
            & $\pmqty{
                 0  & 
                 *  & 
                 * \\ 
                 *  & 
                 *  & 
                 * \\ 
                 *  & 
                 *  & 
                 * \\ 
            }$
            & $\pmqty{
                 *  & 
                 *  & 
                 * \\ 
                 *  & 
                 0  & 
                 * \\ 
                 *  & 
                 *  & 
                 * \\ 
            }$
            & $\pmqty{
                 *  & 
                 *  & 
                 * \\ 
                 *  & 
                 *  & 
                 * \\ 
                 *  & 
                 *  & 
                 0 \\ 
            }$
            \\
            \hline
            \addtocounter{c}{1}
            Type \arabic{c} 
            & $\pmqty{
                 0  & 
                 *  & 
                 * \\ 
                 *  & 
                 *  & 
                 * \\ 
                 *  & 
                 *  & 
                 * \\ 
            }$
            & $\pmqty{
                 *  & 
                 *  & 
                 * \\ 
                 *  & 
                 0  & 
                 * \\ 
                 *  & 
                 *  & 
                 * \\ 
            }$
            & $\pmqty{
                 *  & 
                 *  & 
                 * \\ 
                 *  & 
                 *  & 
                 * \\ 
                 *  & 
                 *  & 
                 * \\ 
            }$
            \\
            \hline
            \addtocounter{c}{1}
            Type \arabic{c} 
            & $\pmqty{
                 0  & 
                 0  & 
                 0 \\ 
                 0  & 
                 *  & 
                 * \\ 
                 0  & 
                 *  & 
                 * \\ 
            }$
            & $\pmqty{
                 0  & 
                 *  & 
                 * \\ 
                 *  & 
                 *  & 
                 * \\ 
                 *  & 
                 *  & 
                 * \\ 
            }$
            & $\pmqty{
                 0  & 
                 *  & 
                 * \\ 
                 *  & 
                 *  & 
                 * \\ 
                 *  & 
                 *  & 
                 * \\ 
            }$
            \\
            \hline
            \addtocounter{c}{1}
            Type \arabic{c} 
            & $\pmqty{
                 0  & 
                 0  & 
                 0 \\ 
                 0  & 
                 *  & 
                 * \\ 
                 0  & 
                 *  & 
                 * \\ 
            }$
            & $\pmqty{
                 0  & 
                 *  & 
                 * \\ 
                 *  & 
                 0  & 
                 * \\ 
                 *  & 
                 *  & 
                 * \\ 
            }$
            & $\pmqty{
                 0  & 
                 *  & 
                 * \\ 
                 *  & 
                 *  & 
                 * \\ 
                 *  & 
                 *  & 
                 * \\ 
            }$
            \\
            \hline
            \addtocounter{c}{1}
            Type \arabic{c} 
            & $\pmqty{
                 0  & 
                 *  & 
                 * \\ 
                 *  & 
                 *  & 
                 * \\ 
                 *  & 
                 *  & 
                 * \\ 
            }$
            & $\pmqty{
                 *  & 
                 *  & 
                 * \\ 
                 *  & 
                 *  & 
                 * \\ 
                 *  & 
                 *  & 
                 * \\ 
            }$
            & $\pmqty{
                 *  & 
                 *  & 
                 * \\ 
                 *  & 
                 *  & 
                 * \\ 
                 *  & 
                 *  & 
                 * \\ 
            }$
            \\
            \hline
            \addtocounter{c}{1}
            Type \arabic{c} 
            & $\pmqty{
                 *  & 
                 *  & 
                 * \\ 
                 *  & 
                 *  & 
                 * \\ 
                 *  & 
                 *  & 
                 * \\ 
            }$
            & $\pmqty{
                 *  & 
                 *  & 
                 * \\ 
                 *  & 
                 *  & 
                 * \\ 
                 *  & 
                 *  & 
                 * \\ 
            }$
            & $\pmqty{
                 *  & 
                 *  & 
                 * \\ 
                 *  & 
                 *  & 
                 * \\ 
                 *  & 
                 *  & 
                 * \\ 
            }$
            \\
            \hline
            \addtocounter{c}{1}
            Type \arabic{c} 
            & $\pmqty{
                 0  & 
                 0  & 
                 0 \\ 
                 0  & 
                 0  & 
                 * \\ 
                 0  & 
                 *  & 
                 * \\ 
            }$
            & $\pmqty{
                 0  & 
                 0  & 
                 * \\ 
                 0  & 
                 0  & 
                 0 \\ 
                 *  & 
                 0  & 
                 * \\ 
            }$
            & $\pmqty{
                 0  & 
                 *  & 
                 * \\ 
                 *  & 
                 0  & 
                 * \\ 
                 *  & 
                 *  & 
                 * \\ 
            }$
            \\
            \hline
            \addtocounter{c}{1}
            Type \arabic{c} 
            & $\pmqty{
                 0  & 
                 0  & 
                 0 \\ 
                 0  & 
                 0  & 
                 * \\ 
                 0  & 
                 *  & 
                 * \\ 
            }$
            & $\pmqty{
                 0  & 
                 0  & 
                 * \\ 
                 0  & 
                 0  & 
                 * \\ 
                 *  & 
                 *  & 
                 * \\ 
            }$
            & $\pmqty{
                 0  & 
                 *  & 
                 * \\ 
                 *  & 
                 *  & 
                 * \\ 
                 *  & 
                 *  & 
                 * \\ 
            }$
            \\
            \hline
            \addtocounter{c}{1}
            Type \arabic{c} 
            & $\pmqty{
                 0  & 
                 0  & 
                 0 \\ 
                 0  & 
                 *  & 
                 * \\ 
                 0  & 
                 *  & 
                 * \\ 
            }$
            & $\pmqty{
                 0  & 
                 *  & 
                 * \\ 
                 *  & 
                 0  & 
                 * \\ 
                 *  & 
                 *  & 
                 * \\ 
            }$
            & $\pmqty{
                 0  & 
                 *  & 
                 * \\ 
                 *  & 
                 *  & 
                 * \\ 
                 *  & 
                 *  & 
                 0 \\ 
            }$
            \\
            \hline
            \addtocounter{c}{1}
            Type \arabic{c} 
            & $\pmqty{
                 0  & 
                 0  & 
                 0 \\ 
                 0  & 
                 0  & 
                 * \\ 
                 0  & 
                 *  & 
                 * \\ 
            }$
            & $\pmqty{
                 0  & 
                 0  & 
                 * \\ 
                 0  & 
                 0  & 
                 * \\ 
                 *  & 
                 *  & 
                 * \\ 
            }$
            & $\pmqty{
                 0  & 
                 *  & 
                 * \\ 
                 *  & 
                 *  & 
                 * \\ 
                 *  & 
                 *  & 
                 0 \\ 
            }$
            \\
            \hline
            \addtocounter{c}{1}
            Type \arabic{c} 
            & $\pmqty{
                 0  & 
                 0  & 
                 0 \\ 
                 0  & 
                 0  & 
                 * \\ 
                 0  & 
                 *  & 
                 * \\ 
            }$
            & $\pmqty{
                 0  & 
                 0  & 
                 * \\ 
                 0  & 
                 0  & 
                 0 \\ 
                 *  & 
                 0  & 
                 * \\ 
            }$
            & $\pmqty{
                 0  & 
                 *  & 
                 * \\ 
                 *  & 
                 0  & 
                 * \\ 
                 *  & 
                 *  & 
                 0 \\ 
            }$
            \\
            \hline
        \end{tabular}
     \end{table}

\subsection{CICYs with $h^{1,1}=2$}
\label{sec:2-moduli}

In this section, we focus on CICYs with $h^{1,1}=2$, which are reviewed in the previous section.
For illustrating models, we analyze the Yukawa textures and mass hierarchies for two generations of matter fields $A^a$ with $a=1,2$. 
We assume that either $A^1$ or $A^2$ plays the role of the Higgs field and develops a vacuum expectation value (VEV). 
As discussed above, four distinct types of Yukawa textures arise in this case. 
To examine their phenomenological implications, we select one example from each type and analyze its mass structure in detail.

\vspace{0.5cm}
$\bullet$~{\bf Type 1 (CICY 7644)}

As an illustrative example of Type 1, we examine CICY 7644.
$(2 \times 2)$ matrices of holomorphic Yukawa couplings are obtained from the intersection numbers as 
\begin{align}
    \kappa_{ab1}=
    \left(
    \begin{array}{cc}
        4 & 12 \\
        12 & 12 \\
    \end{array}
    \right),
\end{align}
when the Higgs field corresponds to $A^1$, and 
\begin{align}
    \kappa_{ab2}=
    \left(
    \begin{array}{cc}
        12 & 12 \\
        12 & 4 \\
    \end{array}
    \right),
\end{align}
when the Higgs field corresponds to $A^2$.
The ratios of the eigenvalues of both matrices are of ${\cal O}(1)$.
However, physical Yukawa couplings can be hierarchical \cite{Ishiguro:2021drk}. 
The K\"ahler metric in Eqs.~(\ref{eq:Kahler-metric}) and (\ref{eq:Kahler-potential}) depends on the K\"ahler moduli and mixes the two generations of matter fields $A^1$ and $A^2$.
We rotate $A^a$ ($a=1,2$) to  $A^{\hat a}$, where 
the K\"ahler metric is diagonal.
Then, we normalize canonically their kinetic terms.
We estimate physical Yukawa couplings $Y_{\hat a \hat b 1}$ 
and $Y_{\hat a \hat b 2}$ in this basis.
These physical Yukawa couplings depend on moduli.
Note that we do not rotate the Higgs basis, which correspond to the original moduli basis.

\begin{figure}[H]
    \centering
    \includegraphics[width=0.7\linewidth]{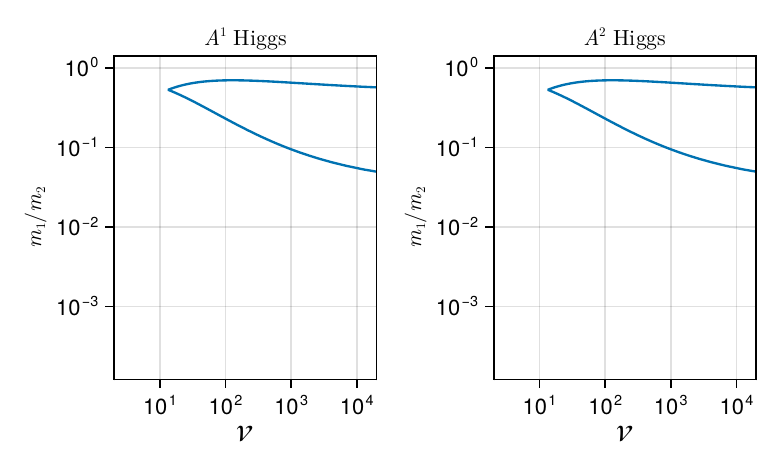}
    \caption{Mass ratios for CICY 7644 as a function of the minimum volume ${\cal V}$ where all real parts of the moduli are greater than or equal to 1. The masses are ordered as $m_1 < m_2$. The left (right) panel shows the ratio $m_1/m_2$ for the first (second) Higgs. The lower curve in the left (right) panel corresponds to the region ${\rm Re}\,T^1 < {\rm Re}\,T^2$ (${\rm Re}\,T^1 >{\rm Re}\,T^2$).}
    \label{fig:h11_2_moduli_2_V_to_ratio_type_1}
\end{figure}

Figure \ref{fig:h11_2_moduli_2_V_to_ratio_type_1} shows the mass ratios $m_1/m_2$, which are obtained from eigenvalue ratios of $Y_{\hat a \hat b 1}$ and $Y_{\hat a \hat b 2}$.
We restrict ourselves to the region of moduli space with ${\rm Re}\,T^a\ge 1$.
As the volume increases, the mass ratio $m_1/m_2$ becomes smaller.
The intersection numbers and holomorphic Yukawa couplings are symmetric under exchange between $T^1$ and $T^2$.
Thus, the curves in the left and right panels of Figure  \ref{fig:h11_2_moduli_2_V_to_ratio_type_1} coincide. 
We need ${\cal V} ={\cal O}(10^{3})$ to realize $m_1/m_2 = {\cal O}(10^{-1})$.

\vspace{0.5cm}
$\bullet$~{\bf Type 2 (CICY 7643)}

As an illustrating example of Type 2, we examine CICY 7643.
The holomorphic Yukawa matrices are obtained from intersection numbers as 
\begin{align}
    \kappa_{ab1}=
    \left(
    \begin{array}{cc}
        0 & 4 \\
        4 & 12 \\
    \end{array}
    \right),
\end{align}
when the Higgs field corresponds to $A^1$, and 
\begin{align}
    \kappa_{ab2}=
    \left(
    \begin{array}{cc}
        4 & 12 \\
        12 & 8 \\
    \end{array}
    \right),
\end{align}
when the Higgs field corresponds to $A^2$.

\begin{figure}[H]
    \centering
    \includegraphics[width=0.7\linewidth]{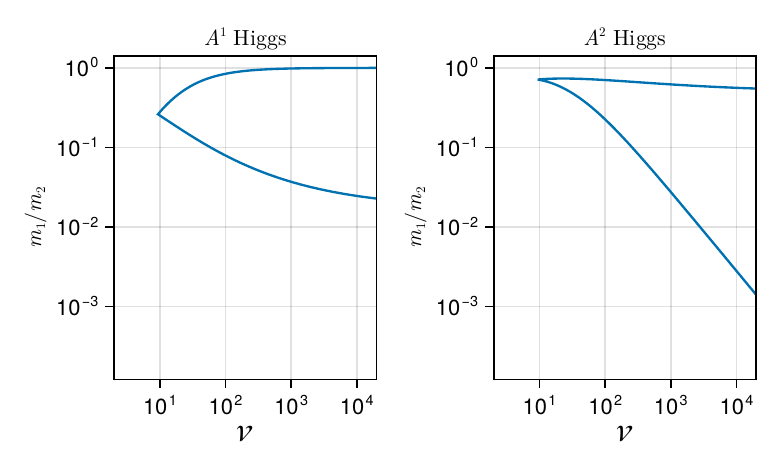}
    \caption{Mass ratios for CICY 7643 as a function of the minimum volume ${\cal V}$ where all real parts of the moduli are greater than or equal to 1. The masses are ordered as $m_1 < m_2$. The left (right) panel shows the ratio $m_1/m_2$ for the first (second) Higgs. The lower curve in the left (right) panel corresponds to the region ${\rm Re}\,T^1 <{\rm Re}\,T^2$ (${\rm Re}\,T^1 > {\rm Re}\,T^2$).}
    \label{fig:h11_2_moduli_2_V_to_ratio_type_2}
\end{figure}

Similar to Type 1, we study physical Yukawa couplings and examine mass ratios.
Figure \ref{fig:h11_2_moduli_2_V_to_ratio_type_2} shows 
mass ratios $m_1/m_2$ depending on the volume.
For ${\cal V} < 10^3$, the case with the Higgs field $A^1$ can lead to smaller ratio $m_1/m_2$, while for ${\cal V} > 10^3$ 
the case with the Higgs field $A^2$ can lead to smaller ratio $m_1/m_2$.

\vspace{0.5cm}
$\bullet$~{\bf Type 3 (CICY 7806)}

As an illustrating example of Type 3, we examine CICY 7806. 
The holomorphic Yukawa matrices are obtained from intersection numbers as

\begin{align}
    \kappa_{ab1}=
    \left(
    \begin{array}{cc}
        0 & 0 \\
        0 & 6 \\
    \end{array}
    \right),
\end{align}
when the Higgs field corresponds to $A^1$, and 
\begin{align}
    \kappa_{ab2}=
    \left(
    \begin{array}{cc}
        0 & 6 \\
        6 & 6 \\
    \end{array}
    \right),
\end{align}
when the Higgs field corresponds to $A^2$.

\begin{figure}[H]
    \centering
    \includegraphics[width=0.7\linewidth]{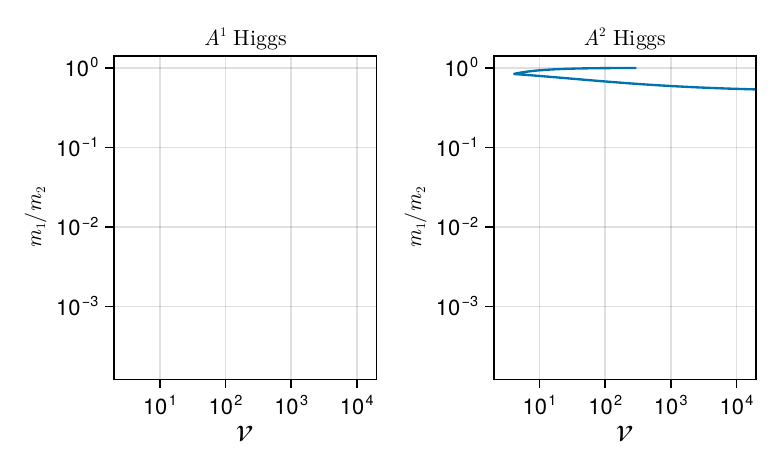}
    \caption{Mass ratios for CICY 7806 as a function of the minimum volume ${\cal V}$ where all real parts of the moduli are greater than or equal to 1. The masses are ordered as $m_1 < m_2$. The left (right) panel shows the ratio $m_1/m_2$ for the first (second) Higgs. The lower curve in the right panel corresponds to the region ${\rm Re}\,T^1 <{\rm Re}\,T^2$.}
    \label{fig:h11_2_moduli_2_V_to_ratio_type_3}
\end{figure}

Figure \ref{fig:h11_2_moduli_2_V_to_ratio_type_3} shows mass ratios depending on the volume.
In this example, the $(2 \times 2)$ matrix $Y_{\hat a \hat b 1}$  as well as $\kappa_{ab1}$ has rank 1.
Thus, we have $m_1=0$ for any moduli values when the Higgs field correspond to $A^1$.
When the Higgs field corresponds to $A^2$, the mass ratio $m_1/m_2$ is not suppressed compared with other examples.

\vspace{0.5cm}
$\bullet$~{\bf Type 4 (CICY 7884)}

As an example of Type 4, we examine CICY 7884.
The holomorphic Yukawa matrices are obtained from the intersecting numbers as 

\begin{align}
    \kappa_{ab1}=
    \left(
    \begin{array}{cc}
        0 & 3 \\
        3 & 3 \\
    \end{array}
    \right),
\end{align}
when the Higgs field corresponds to 
$A^1$, and 
\begin{align}
    \kappa_{ab2}=
    \left(
    \begin{array}{cc}
        3 & 3 \\
        3 & 0 \\
    \end{array}
    \right),
\end{align}
when the Higgs field corresponds to $A^2$.

\begin{figure}[H]
    \centering
    \includegraphics[width=0.7\linewidth]{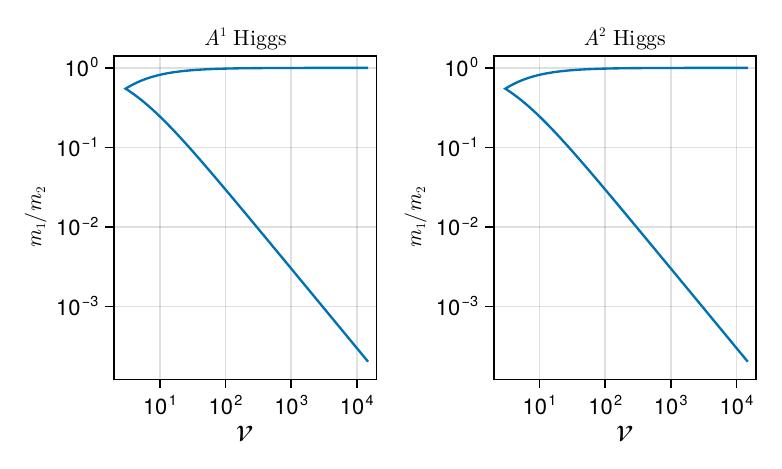}
    \caption{Mass ratios for CICY 7884 as a function of the minimum volume ${\cal V}$ where all real parts of the moduli are greater than or equal to 1. The masses are ordered as $m_1 < m_2$. The left (right) panel shows the ratio $m_1/m_2$ for the first (second) Higgs.  The lower curve in the left (right) panel corresponds to the region ${\rm Re}\,T^1 < {\rm Re}\,T^2$ (${\rm Re}\,T^1 > {\rm Re}\,T^2$).}
    \label{fig:h11_2_moduli_2_V_to_ratio_type_4}
\end{figure}

Figure \ref{fig:h11_2_moduli_2_V_to_ratio_type_4} shows mass ratios depending on the volume.
The intersecting numbers and holomorphic Yukawa couplings are symmetric under exchange between $T^1$ and $T^2$.
Thus, the curves in left and right panels of Figure \ref{fig:h11_2_moduli_2_V_to_ratio_type_4} coincide. 
This example leads to most suppressed mass ratio $m_1/m_2$ among four examples.
For example, we can realize $m_1/m_2={\cal O}(10^{-2})$ for  $V={\cal O}(10^{2.5})$.

\subsection{CICYs with $h^{1,1}=3$}
\label{sec:3-moduli}

Similar to the previous section, here  we study Yukawa textures and mass ratios for CICYs with $h^{1,1}=3$.
There are three generations of matter fields $A^a$ with $a=1,2,3$.
We assume that one of the $A^a$ ($a=1,2,3$) plays the role of the Higgs field and develops its VEV.
In this case, eleven distinct types arise. 
To investigate their phenomenological aspects, we select one representative example from each type and analyze it in detail. 
In particular, we present three types, namely Type 1, Type 6 and Type 10. The other cases are shown in Appendix~\ref{app:h11_3}.

\vspace{0.5cm}
$\bullet$~{\bf Type 1 (CICY 5299)}

As an illustrative example of Type 1, we examine CICY 5299.
The holomorphic Yukawa matrices are obtained from the intersection numbers as 
\begin{align}
\kappa_{ab1}=
\begin{pmatrix}
0 & 2 & 4 \\
2 & 4 & 9 \\
4 & 9 & 2 \\
\end{pmatrix}
,
\end{align}
when the Higgs field corresponds to $A^1$,
\begin{align}
\kappa_{ab2}=
\begin{pmatrix}
2 & 4 & 9 \\
4 & 0 & 2 \\
9 & 2 & 4 \\
\end{pmatrix}
,
\end{align}
when the Higgs field corresponds to $A^2$, and 
\begin{align}
\kappa_{ab3}=
\begin{pmatrix}
4 & 9 & 2 \\
9 & 2 & 4 \\
2 & 4 & 0 \\
\end{pmatrix}
,
\end{align}
when the Higgs field corresponds to $A^3$.

\begin{figure}[H]
	\centering
	\includegraphics[width=0.7\linewidth]{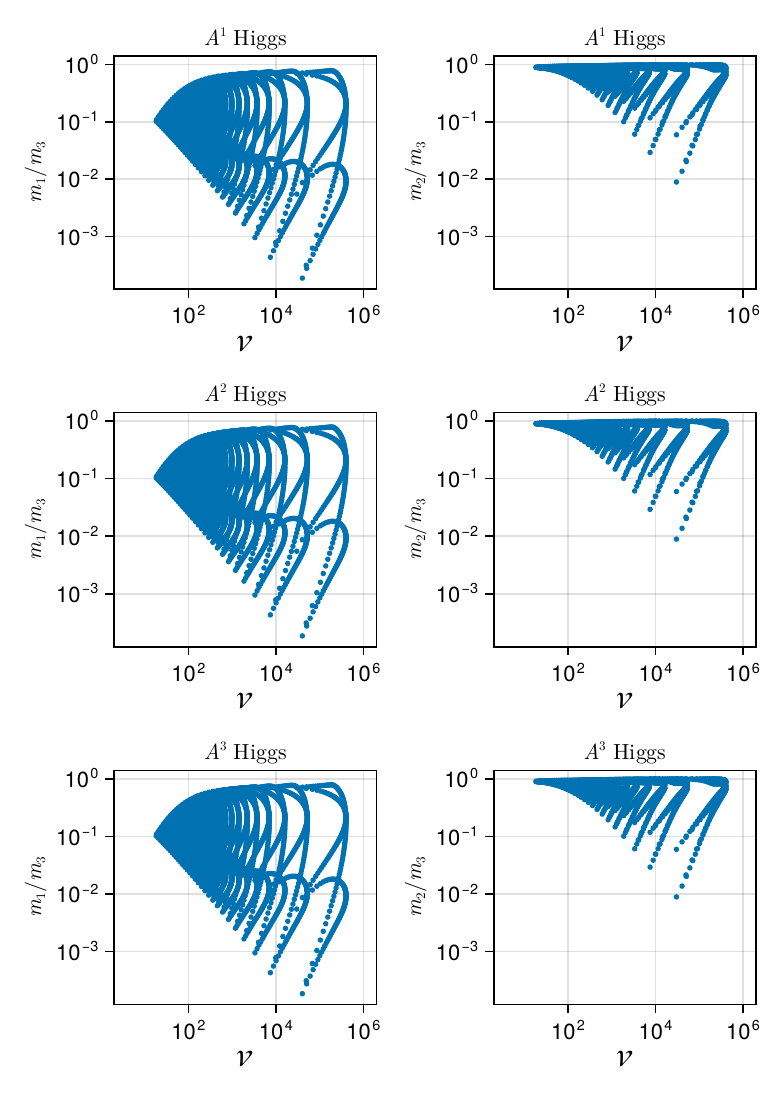}
	\caption{Mass ratios for CICY 5299 as a function of the minimum volume ${\cal V}$ where all real parts of the moduli are greater than or equal to 1. The masses are ordered as $m_1 < m_2 < m_3$. The left (right) panel shows the scatter for $m_1/m_3$ ($m_2/m_3$).}
	\label{fig:h11_3_moduli_3_V_to_ratio_type_1}
\end{figure}

Figure \ref{fig:h11_3_moduli_3_V_to_ratio_type_1} shows the mass ratios $m_1/m_3$ and $m_2/m_3$, which are obtained from eigenvalue ratios of $Y_{\hat a \hat b 1}$, $Y_{\hat a \hat b 2}$ and $Y_{\hat a \hat b 3}$. The points correspond to discrete variations of the moduli ratio $T^1:T^2:T^3$, subject to the restriction ${\rm Re}\,T^a\ge 1$.

\vspace{0.5cm}
$\bullet$~{\bf Type 6 (CICY 7242)}

As an illustrative example of Type 6, we examine CICY 7242.
The holomorphic Yukawa matrices are obtained from the intersection numbers as 
\begin{align}
\kappa_{ab1}=
\begin{pmatrix}
2 & 6 & 6 \\
6 & 6 & 10 \\
6 & 10 & 6 \\
\end{pmatrix}
,
\end{align}
when the Higgs field corresponds to $A^1$,
\begin{align}
\kappa_{ab2}=
\begin{pmatrix}
6 & 6 & 10 \\
6 & 2 & 6 \\
10 & 6 & 6 \\
\end{pmatrix}
,
\end{align}
when the Higgs field corresponds to $A^2$, and 
\begin{align}
\kappa_{ab3}=
\begin{pmatrix}
6 & 10 & 6 \\
10 & 6 & 6 \\
6 & 6 & 2 \\
\end{pmatrix}
,
\end{align}
when the Higgs field corresponds to $A^3$.

\begin{figure}[H]
	\centering
	\includegraphics[width=0.7\linewidth]{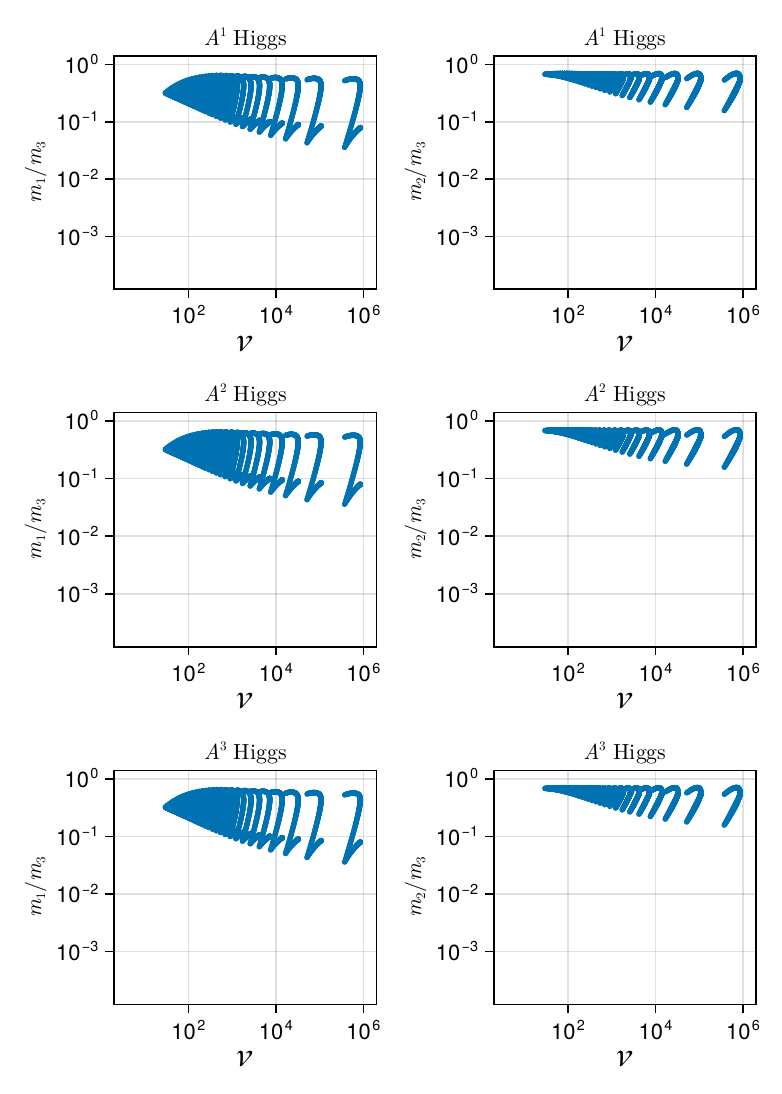}
	\caption{Mass ratios for CICY 7242 as a function of the minimum volume ${\cal V}$ where all real parts of the moduli are greater than or equal to 1. The masses are ordered as $m_1 < m_2 < m_3$. The left (right) panel shows the scatter for $m_1/m_3$ ($m_2/m_3$).}
	\label{fig:h11_3_moduli_3_V_to_ratio_type_6}
\end{figure}

Figure \ref{fig:h11_3_moduli_3_V_to_ratio_type_6} shows the mass ratios $m_1/m_3$ and $m_2/m_3$, which are obtained from eigenvalue ratios of $Y_{\hat a \hat b 1}$, $Y_{\hat a \hat b 2}$ and $Y_{\hat a \hat b 3}$. The points correspond to discrete variations of the moduli ratio $T^1:T^2:T^3$, subject to the restriction ${\rm Re}\,T^a\ge 1$.

\vspace{0.5cm}
$\bullet$~{\bf Type 10 (CICY 7875)}

As an illustrative example of Type 10, we examine CICY 7875.
The holomorphic Yukawa matrices are obtained from the intersection numbers as 
\begin{align}
\kappa_{ab1}=
\begin{pmatrix}
0 & 0 & 0 \\
0 & 0 & 3 \\
0 & 3 & 2 \\
\end{pmatrix}
,
\end{align}
when the Higgs field corresponds to $A^1$,
\begin{align}
\kappa_{ab2}=
\begin{pmatrix}
0 & 0 & 3 \\
0 & 0 & 3 \\
3 & 3 & 3 \\
\end{pmatrix}
,
\end{align}
when the Higgs field corresponds to $A^2$, and 
\begin{align}
\kappa_{ab3}=
\begin{pmatrix}
0 & 3 & 2 \\
3 & 3 & 3 \\
2 & 3 & 0 \\
\end{pmatrix}
,
\end{align}
when the Higgs field corresponds to $A^3$.

\begin{figure}[H]
	\centering
	\includegraphics[width=0.7\linewidth]{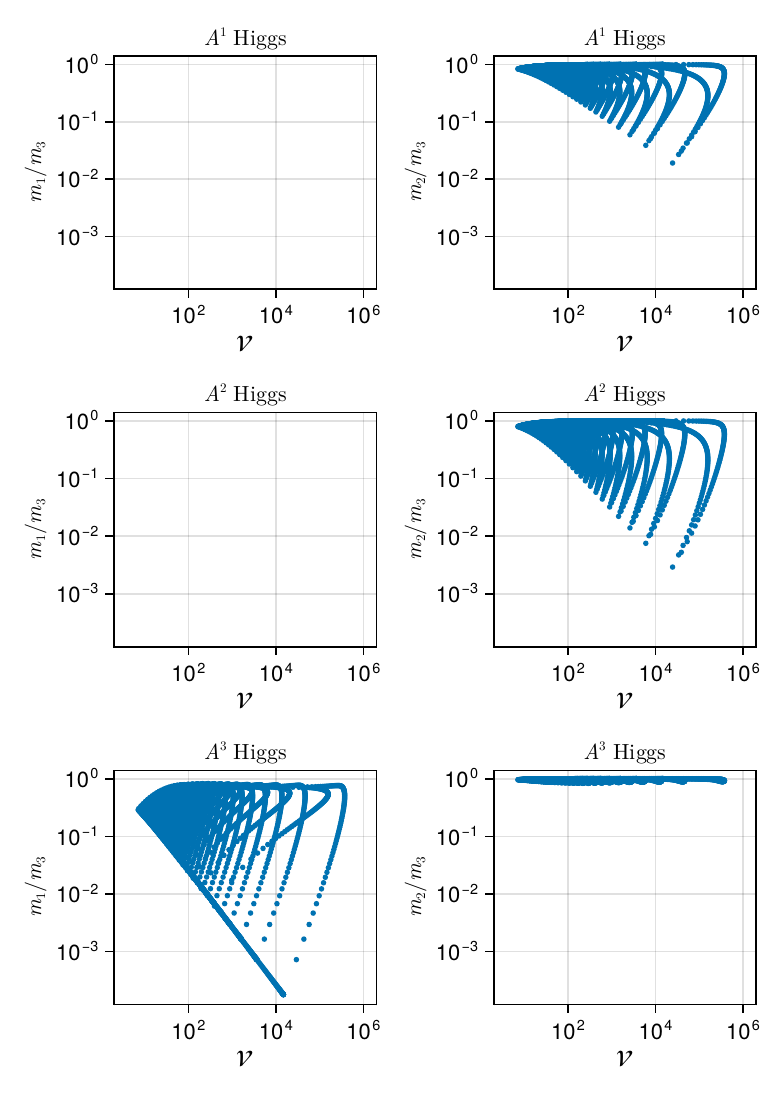}
	\caption{Mass ratios for CICY 7875 as a function of the minimum volume ${\cal V}$ where all real parts of the moduli are greater than or equal to 1. The masses are ordered as $m_1 < m_2 < m_3$. The left (right) panel shows the scatter for $m_1/m_3$ ($m_2/m_3$).}
	\label{fig:h11_3_moduli_3_V_to_ratio_type_10}
\end{figure}

Figure \ref{fig:h11_3_moduli_3_V_to_ratio_type_10} shows the mass ratios $m_1/m_3$ and $m_2/m_3$, which are obtained from eigenvalue ratios of $Y_{\hat a \hat b 1}$, $Y_{\hat a \hat b 2}$ and $Y_{\hat a \hat b 3}$. The points correspond to discrete variations of the moduli ratio $T^1:T^2:T^3$, subject to the restriction ${\rm Re}\,T^a\ge 1$.
In this example, the $(3\times 3)$ matrices  $Y_{\hat a \hat b 1}$ and $Y_{\hat a \hat b 2}$ have rank 2.
When the Higgs field corresponds to $A^1$ and $A^2$, we have $m_1=0$ for any moduli values.

\subsection{Phenomenological implications}
\label{sec:pheno}

The results for CICYs with $h^{1,1}=3$ in the previous section show that mass hierarchies, in particular small values of $m_1/m_3$, can be realized when the volume $\cal V$ is large.
For ${\cal V}=10^2$, we obtain $m_1/m_3=10^{-1}-10^{-2}$.
For ${\cal V}=10^3$, even smaller values such as $m_1/m_3=10^{-3}$ or below can be achieved.

Quark and lepton mass ratios are estimated e.g. at $10^{16}$ GeV as \cite{Antusch:2025fpm}
\begin{align}
    \frac{m_c}{m_t}\sim 3\times 10^{-3}, \quad \frac{m_u}{m_t} \sim 6 \times 10^{-6}, \notag \\
    \frac{m_s}{m_b}\sim 2\times 10^{-2}, \quad \frac{m_d}{m_b} \sim 1 \times 10^{-3}, \notag \\
        \frac{m_\mu}{m_\tau}\sim 6\times 10^{-2}, \quad \frac{m_e}{m_\tau} \sim 3 \times 10^{-4}. 
\end{align}
In order to realize these large mass hierarchies, we need a large CY volume ${\cal V}$. 
However, ${\cal V}$ is related to the four-dimensional gauge coupling via Eq.~(\ref{eq:g-V}). 
The gauge coupling 
at the GUT scale is estimated as  $\alpha^{-1}(M_{\rm GUT}) \simeq 25$ with $M_{\rm GUT} \simeq 2\times 10^{16}$\,GeV in the minimal supersymmetric Standard Model, 
while we estimate as 
$\alpha^{-1}(M_{\rm GUT}) \sim 50$ in the Standard Model. 
Requiring the weak string coupling $g_s \leq 1$ leads to the constraint on the volume as ${\cal V} \lesssim 25$ or $50$, depending on the running coupling behavior.
Under this constraint, we may realize smaller hierarchy like $m_1/m_3 \gtrsim 10^{-1}-10^{-2}$ from the results in the previous section. 
If additional effects from lighter gauge bosons are present,\footnote{The correction $\Delta \alpha^{-1}$ is estimated as 
$\Delta \alpha^{-1} 
=b\ln M_{\rm GUT}/M_V$, where $b$ is the one-loop beta function coefficient and $M_V$ denotes a vector boson mass lighter than $M_{\rm GUT}$.} we may obtain larger value of $\alpha^{-1}(M_{\rm GUT})$, allowing for larger ${\cal V}$.

So far, we have assumed that the (light) Higgs field, which develops its VEV, corresponds to one of the $A^a$ in the original moduli basis. 
However, the light Higgs direction may correspond to the mixed direction of the original moduli basis, $\sum_a c_aA^a$ in general. 
Indeed, generic string compactifications typically contain many candidates for the Higgs fields. 
That is the multi-Higgs models at the compactification scale.
The light Higgs field would be determined by their mass terms, which may be generated by VEVs of other fields and/or non-perturbative effects.
The precise structure depends on the details of mass generation mechanism.
Such a discussion on mass generation scenarios is beyond our scope. 
Here we parametrize the light Higgs direction and demonstrate interesting results.

Suppose that the light Higgs direction, which develop its VEV, is given by $A^1 + \varepsilon A^3$ in the CICY 7806 model of type 3 with $h^{1,1}=2$. 
Then the holomorphic $(2\times 2)$ Yukawa matrix takes the form: 
\begin{align}
    \kappa_{ab}\sim 6\begin{pmatrix}
        0 &\varepsilon \\
        \varepsilon & 1
    \end{pmatrix}.
\end{align}
This structure leads to a mass hierarchy as well as a mixing angle controlled by $\varepsilon$, similar to the down sector of the Weinberg texture \cite{Weinberg:1977hb}.

\begin{figure}[H]
	\centering
	\includegraphics[width=0.7\linewidth]{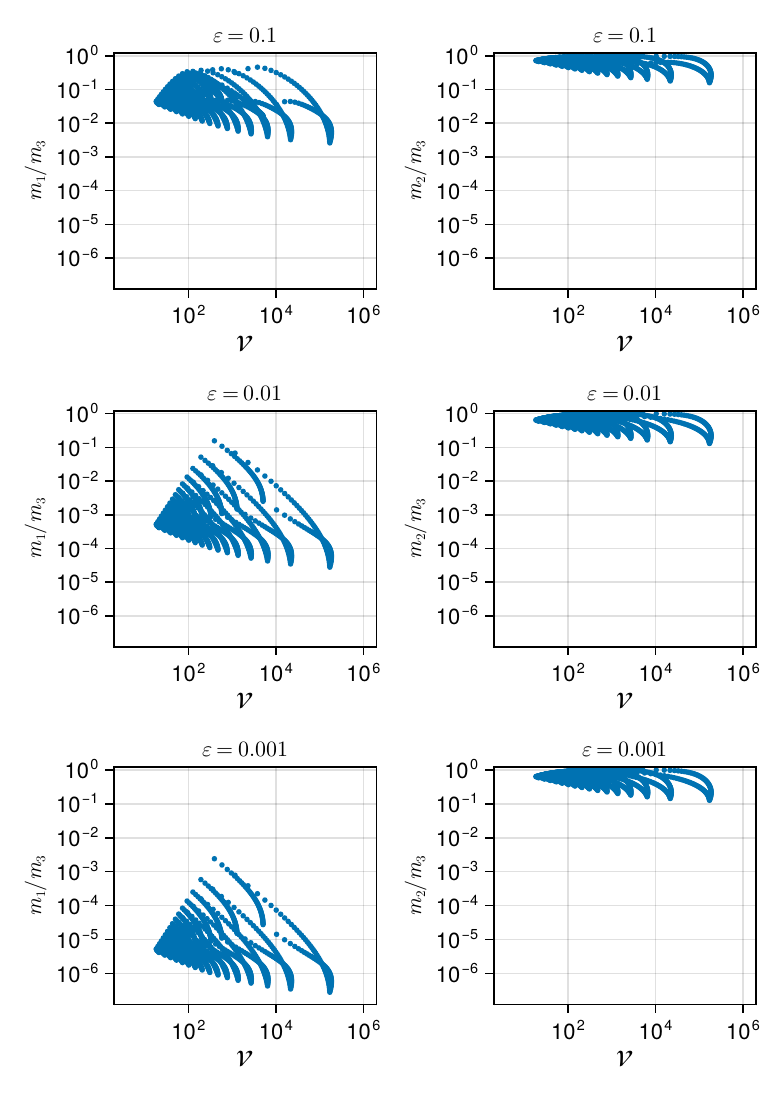}
	\caption{Mass ratios for CICY 7465 as a function of the minimum volume ${\cal V}$ when the Higgs field is assigned to $A^1 + \varepsilon A^3$ ($\varepsilon = 0.1,\, 0.01,\, 0.001$), where all real parts of the moduli are greater than or equal to 1. The masses are ordered as $m_1 < m_2 < m_3$. The left (right) panel shows the scatter for $m_1/m_3$ ($m_2/m_3$).}
	\label{fig:h11_3_moduli_3_V_to_ratio_type_8_with_epsilon}
\end{figure}

Similarly, suppose that the light Higgs direction, which develops its VEV, is written by $A^1 + \varepsilon A^3$ in CICY 7465 of type 10 with $h^{1,1}=3$. 
We set $\varepsilon = 0.1, 0.01, 0.001$. 
Figure \ref{fig:h11_3_moduli_3_V_to_ratio_type_8_with_epsilon} shows the mass ratios. 
The mass ratio $m_1/m_3$ can be small depending on $\varepsilon$, even for ${\cal V} \lesssim 50$.
Note that the mass matrix has rank 2 along the $A^1$ direction, 
while the mass matrix has rank 3 along the $A^3$ direction. 
This interplay leads to the hierarchical structure.
Also, we can find the direction leading to the rank-2 mass matrix in other CICY models with $h^{1,1}=3$.
However, we can not find rank-1 ($3\times 3$) mass matrices in CICY models with $h^{1,1}=3$.

\begin{figure}[H]
	\centering
	\includegraphics[width=0.7\linewidth]{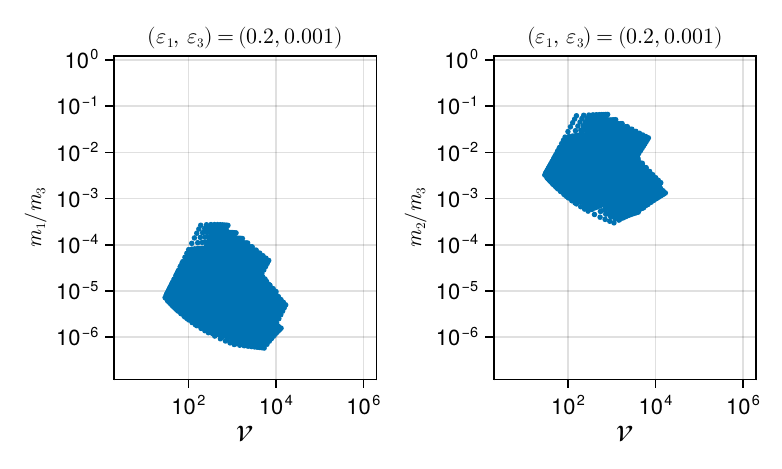}
	\caption{Mass ratios for CICY 5245 as a function of the minimum volume ${\cal V}$ when the Higgs field is assigned to $A^0 + 0.2 \cdot A^1 + 0.001 \cdot A^3$ ($A^0 = 3A^1 + 6A^2 - 4A^3 + 2A^4$), where all real parts of the moduli are greater than or equal to 1. The three fermion generations are assigned to $A^1, A^2, A^4$. The masses are ordered as $m_1 < m_2 < m_3$. The left (right) panel shows the scatter for $m_1/m_3$ ($m_2/m_3$).}
	\label{fig:h11_4_moduli_3_V_to_ratio_type_3_with_epsilon_uct}
\end{figure}

\begin{figure}[H]
	\centering
	\includegraphics[width=0.7\linewidth]{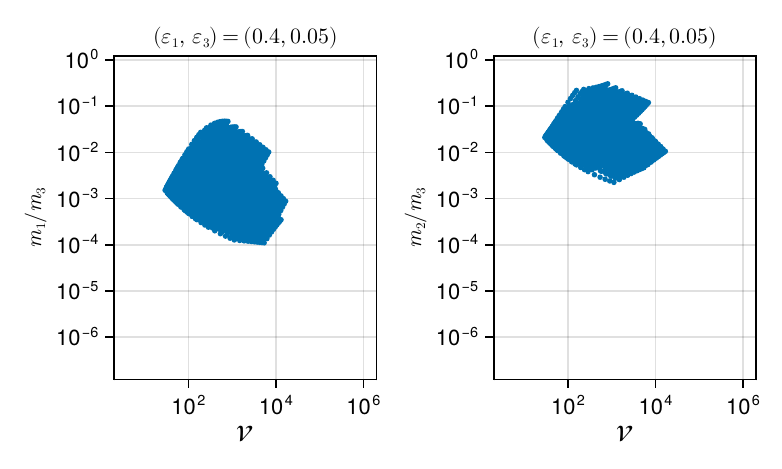}
	\caption{Mass ratios for CICY 5245 as a function of the minimum volume ${\cal V}$ when the Higgs field is assigned to $A^0 + 0.4 \cdot A^1 + 0.05 \cdot A^3$ ($A^0 = 3A^1 + 6A^2 - 4A^3 + 2A^4$), where all real parts of the moduli are greater than or equal to 1. The three fermion generations are assigned to $A^1, A^2, A^4$. The masses are ordered as $m_1 < m_2 < m_3$. The left (right) panel shows the scatter for $m_1/m_3$ ($m_2/m_3$).}
	\label{fig:h11_4_moduli_3_V_to_ratio_type_3_with_epsilon_dsb}
\end{figure}

\begin{figure}[H]
	\centering
	\includegraphics[width=0.7\linewidth]{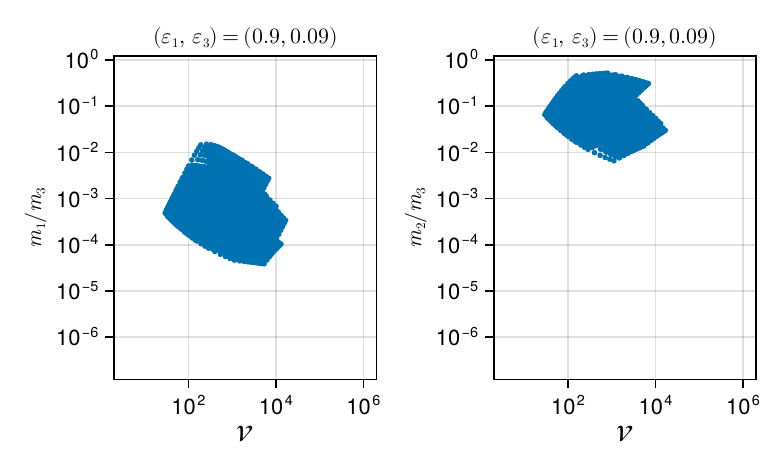}
	\caption{Mass ratios for CICY 5245 as a function of the minimum volume ${\cal V}$ when the Higgs field is assigned to $A^0 + 0.9 \cdot A^1 + 0.09 \cdot A^3$ ($A^0 = 3A^1 + 6A^2 - 4A^3 + 2A^4$), where all real parts of the moduli are greater than or equal to 1. The three fermion generations are assigned to $A^1, A^2, A^4$. The masses are ordered as $m_1 < m_2 < m_3$. The left (right) panel shows the scatter for $m_1/m_3$ ($m_2/m_3$).}
	\label{fig:h11_4_moduli_3_V_to_ratio_type_3_with_epsilon_lepton}
\end{figure}

We extend the above analysis to CICY models with $h^{1,1}=4$. 
Their coupling selection rules are classified in Ref.~\cite{Dong:2025pah}. 
We use CICY 5245 as an example.
We assume that three  $A^1$,$A^2$, and $A^4$ among four $A^a$ correspond to three generations. 
Then, the $(3\times 3)$ mass matrix becomes rank-1, i.e., $m_1=m_2=0$ if the light Higgs direction, which develops its VEV, corresponds to 
\begin{align}
    A^0=3A^1+6A^2-4A^3+2A^4.
\end{align}
A small deviation from this direction may generate large hierarchies among $m_1,m_2$ and $m_3$.
For example, we consider the deviation like $A^0+\varepsilon_1A^1+\varepsilon_3A^3$. 
Figures \ref{fig:h11_4_moduli_3_V_to_ratio_type_3_with_epsilon_uct}, \ref{fig:h11_4_moduli_3_V_to_ratio_type_3_with_epsilon_dsb} and \ref{fig:h11_4_moduli_3_V_to_ratio_type_3_with_epsilon_lepton} show 
the mass ratios $m_1/m_3$ and $m_2/m_3$ when 
$(\varepsilon_1,\varepsilon_3)=(0.2,0.001)$, $(0.4,0.05)$, and $(0.9,0.09)$.
In the limit that $\varepsilon_1, \varepsilon_3 \to 0$, two mass eigenvalues $m_1$ and $m_2$ vanish and a $U(2)$ flavor symmetry emerges. 
Such $U(2)$ flavor symmetry and its small violation are phenomenologically interesting in particle physics, e.g., in controlling flavor-dependent operators in the Standard Model effective field theory \cite{Barbieri:2011ci,Blankenburg:2012nx,Barbieri:2012uh,Fuentes-Martin:2019mun}. 
This scenario provides us with a concrete realization of $U(2)$ flavor symmetry scenario.
Furthermore, higher-dimensional operators are also governed by the prepotential of moduli fields~\cite{Bershadsky:1993cx}, and this observation supports the hypothesis of minimal flavor
violation in string-derived low energy effective field theory~\cite{Kobayashi:2021uam}.\footnote{See also Refs.~\cite{Kobayashi:2021pav,Kobayashi:2022jvy,Kang:2026qgi}.}

Moreover, let us assume that the light Higgs directions in the up-sector and dow-sector are different from each other, analogous to the minimal supersymmetric Standard Model:
$A^0+\varepsilon_{u1}A^1+\varepsilon_{u3}A^3$ for the up-sector and 
$A^0+\varepsilon_{d1}A^1+\varepsilon_{d3}A^3$ for the up-sector. 
Table \ref{tab:h11_4_moduli_3_values} shows representative examples of mass ratios and mixing angles, derived from specific values of the moduli $T^a$ and parameters $\varepsilon_{u1}, \varepsilon_{u3}, \varepsilon_{d1}, \varepsilon_{d3}$. 
They realize the orders of experimental values, although we do not fully reproduce all of experimental data for mass hierarchies and mixing angles. 
Hence, this direction to study is interesting and has a potential to realize experimental data. 
We would study more systematically other CICY models elsewhere.

As emphasized above, Yukawa textures, in particular rank-reduced matrices and small perturbations around them, play a crucial role in these types of models.\footnote{See Ref.~\cite{Hoshiya:2022qvr} for a similar scenario in magnetized orbifold models.}
At any rate, the light Higgs direction is determined by the mass matrix of Higgs fields. 
If they depend on moduli, we may need fine-tuning to realize the hierarchical structures discussed above. 
One of the interesting possibilities is that moduli are trapped dynamically at special points where additional massless modes appear \cite{Kofman:2004yc,Kikuchi:2023uqo}. 
In such a scenario, the moduli may be trapped around the point, where the rank of the mass matrix is reduced, although we need a certain mechanism to stabilize moduli around the point with an enhanced symmetry.

\begin{table}[H]
    \caption{Mass ratios and mixing angles.}
    \label{tab:h11_4_moduli_3_values}
\begin{tabular}{|c|r|c|c|c|c|}    
        \hline
        $(T^1,T^2,T^3,T^4)$ & ${\cal V}$ & $\begin{pmatrix} \varepsilon_{u1} \\ \varepsilon_{u3} \end{pmatrix}$ & $\begin{pmatrix} \varepsilon_{d1} \\ \varepsilon_{d3} \end{pmatrix}$ & $\begin{pmatrix} m_u/m_t \\ m_c/m_t \\ m_d/m_b \\ m_s/m_b \end{pmatrix}$ & $\begin{pmatrix} |V_{us}| \\ |V_{cb}| \\ |V_{ub}| \end{pmatrix}$ \\
        \hline
        \hline
        $(1,4,1,1)$ & 71 & $\begin{pmatrix} 0.10 \\ 3.2 \times 10^{-4} \end{pmatrix}$ & $\begin{pmatrix} 0.18 \\ 7.9 \times 10^{-3} \end{pmatrix}$ & $\begin{pmatrix} 5 \times 10^{-6} \\ 4 \times 10^{-3} \\ 6 \times 10^{-4} \\ 2 \times 10^{-2} \end{pmatrix}$ & $\begin{pmatrix} 0.20 \\ 0.052 \\ 0.006 \end{pmatrix}$ \\
        \hline
        $(1,3,1,1)$ & 57 & $\begin{pmatrix} 0.11 \\ 3.2 \times 10^{-4} \end{pmatrix}$ & $\begin{pmatrix} 0.18 \\ 7.9 \times 10^{-3} \end{pmatrix}$ & $\begin{pmatrix} 4 \times 10^{-6} \\ 3 \times 10^{-3} \\ 6 \times 10^{-4} \\ 1 \times 10^{-2} \end{pmatrix}$ & $\begin{pmatrix} 0.20 \\ 0.037 \\ 0.005 \end{pmatrix}$ \\
        \hline
        $(1,1.25,1,1)$ & 33 & $\begin{pmatrix} 0.16 \\ 5.0 \times 10^{-4} \end{pmatrix}$ & $\begin{pmatrix} 0.28 \\ 0.016 \end{pmatrix}$ & $\begin{pmatrix} 3 \times 10^{-6} \\ 3 \times 10^{-3} \\ 6 \times 10^{-4} \\ 1 \times 10^{-2} \end{pmatrix}$ & $\begin{pmatrix} 0.20 \\ 0.038 \\ 0.008 \end{pmatrix}$ \\
        \hline
        $(1.25,1.25,1,1)$ & 38 & $\begin{pmatrix} 0.18 \\ 1.0 \times 10^{-4} \end{pmatrix}$ & $\begin{pmatrix} 0.28 \\ 0.013 \end{pmatrix}$ & $\begin{pmatrix} 1 \times 10^{-7} \\ 3 \times 10^{-3} \\ 6 \times 10^{-4} \\ 1 \times 10^{-2} \end{pmatrix}$ & $\begin{pmatrix} 0.23 \\ 0.032 \\ 0.006 \end{pmatrix}$ \\
        \hline
        $(1,1,1,1)$ & 29 & $\begin{pmatrix} 0.22 \\ 0.010 \end{pmatrix}$ & $\begin{pmatrix} 0.11 \\ 0.040 \end{pmatrix}$ & $\begin{pmatrix} 4 \times 10^{-4} \\ 5 \times 10^{-3} \\ 4 \times 10^{-3} \\ 1 \times 10^{-2} \end{pmatrix}$ & $\begin{pmatrix} 0.30 \\ 0.027 \\ 0.006 \end{pmatrix}$ \\
        \hline
    \end{tabular}
\end{table}

\section{Conclusions}
\label{sec:con}

In this paper, we have clarified the Yukawa textures of matter fields corresponding to moduli fields in the context of heterotic string theory on CY threefolds with standard embedding. 
In particular, we focused on CICYs with a small number of moduli. 
Since the selection rules for matter fields are governed by those of the moduli fields, the topological structure of the CY threefold leads to novel Yukawa structures that cannot be obtained from group-theoretical symmetries. 

By analyzing the Yukawa textures of matter fields on CICYs with $h^{1,1}=2$ in section~\ref{sec:2-moduli} and $h^{1,1}=3$ in section~\ref{sec:3-moduli}, we found that hierarchical structures cannot be realized in the moduli basis. 
This is because the CY volume is related to the magnitude of four-dimensional gauge coupling at the compactification scale, 
preventing us from taking the large volume regime. 
To overcome this problem, we analyzed a more generic region of the moduli space of multi-Higgs fields, including the Standard Model Higgs field as the lightest mode. 
Remarkably,  in the case of CICYs with $h^{1,1}=2$, we can realize the so-called Weinberg texture, which successfully reproduces the Cabibbo angle by the mass ratio between the down and strange quarks. 
Furthermore, when the Higgs field in the Standard Model is aligned with special loci in the moduli space, the rank of the fermion mass matrix is reduced to two for CICYs with $h^{1,1}=3$ and to one for certain CICYs with $h^{1,1}=4$. 
Hence, in the rank-1 case realized in some CICYs with $h^{1,1}=4$, a $U(2)$ flavor symmetry emerges at these special loci in the moduli space of multi-Higgs fields. 
It is well known that such a $U(2)$ flavor symmetry plays an important role in controlling higher-dimensional operators in the Standard Model effective field theory. 
Our results provide a string-theoretic framework that supports this scenario within string-derived low energy effective field theory. 

When we consider small perturbations around the symmetry-enhanced loci, it is expected to generate hierarchical fermion masses. 
Indeed, our numerical analysis shows that CY volumes consistent with phenomenologically viable gauge couplings can reproduce the observed masses and mixing angles in the quark sector. 
It would be fascinating to figure out the stabilization mechanism of the Higgs fields as well as moduli fields that leads to realistic patterns of fermion masses and mixings. We will leave this important issue for future work.

\acknowledgments

This work was supported by JSPS KAKENHI Grant Numbers JP23K03375 (T.K.) and JP25H01539 (H.O.).

\appendix

\section{$h^{1,1}=3$}
\label{app:h11_3}

In this section, we summarize the Yukawa textures of matter fields on CICYs except for Type 1, Type 6 and Type 10. 

\vspace{0.5cm}
$\bullet$~{\bf Type 2 (CICY 6220)}

As an illustrative example of Type 2, we examine CICY 6220.
The holomorphic Yukawa matrices are obtained from the intersection numbers as 
\begin{align}
\kappa_{ab1}=
\begin{pmatrix}
0 & 3 & 2 \\
3 & 3 & 9 \\
2 & 9 & 6 \\
\end{pmatrix}
,
\end{align}
when the Higgs field corresponds to $A^1$,
\begin{align}
\kappa_{ab2}=
\begin{pmatrix}
3 & 3 & 9 \\
3 & 0 & 2 \\
9 & 2 & 6 \\
\end{pmatrix}
,
\end{align}
when the Higgs field corresponds to $A^2$, and 
\begin{align}
\kappa_{ab3}=
\begin{pmatrix}
2 & 9 & 6 \\
9 & 2 & 6 \\
6 & 6 & 4 \\
\end{pmatrix}
,
\end{align}
when the Higgs field corresponds to $A^3$.

\begin{figure}[H]
	\centering
	\includegraphics[width=0.7\linewidth]{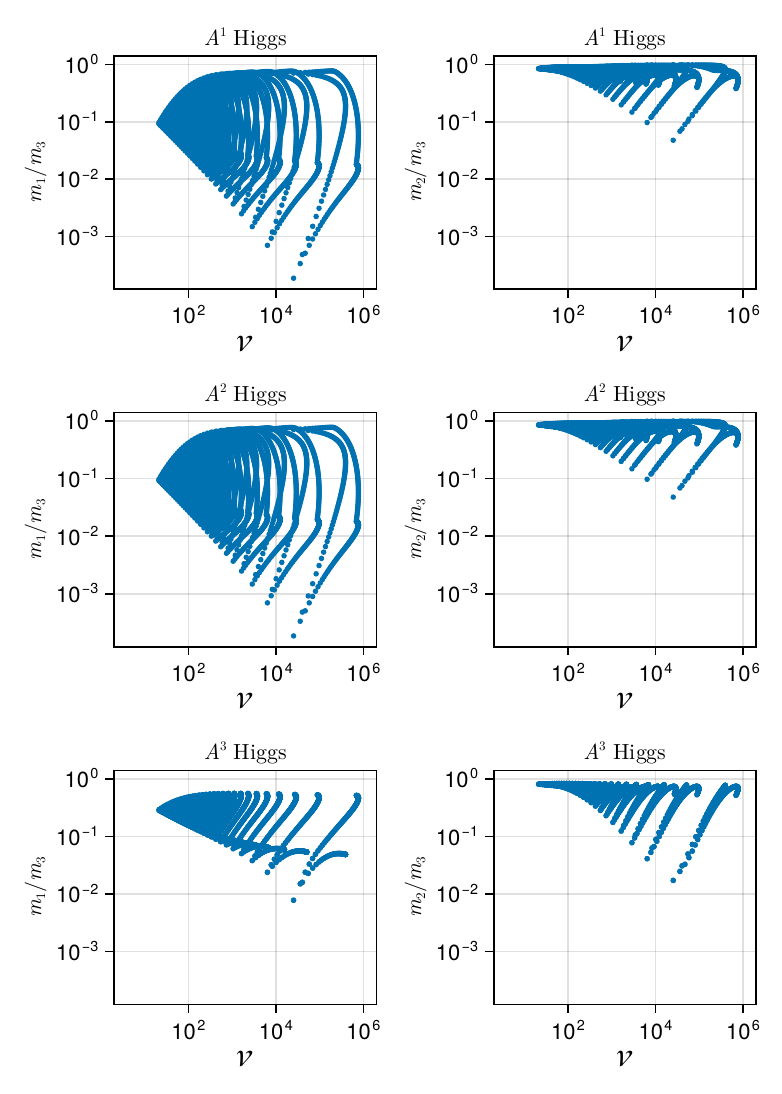}
	\caption{Mass ratios for CICY 6220 as a function of the minimum volume ${\cal V}$ where all real parts of the moduli are greater than or equal to 1. The masses are ordered as $m_1 < m_2 < m_3$. The left (right) panel shows the scatter for $m_1/m_3$ ($m_2/m_3$).}
	\label{fig:h11_3_moduli_3_V_to_ratio_type_2}
\end{figure}

Figure \ref{fig:h11_3_moduli_3_V_to_ratio_type_2} shows the mass ratios $m_1/m_3$ and $m_2/m_3$, which are obtained from eigenvalue ratios of $Y_{\hat a \hat b 1}$, $Y_{\hat a \hat b 2}$ and $Y_{\hat a \hat b 3}$. The points correspond to discrete variations of the moduli ratio $T^1:T^2:T^3$, subject to the restriction ${\rm Re}\,T^a\ge 1$.

\vspace{0.5cm}
$\bullet$~{\bf Type 3 (CICY 6771)}

As an illustrative example of Type 3, we examine CICY 6771.
The holomorphic Yukawa matrices are obtained from the intersection numbers as 
\begin{align}
\kappa_{ab1}=
\begin{pmatrix}
0 & 0 & 0 \\
0 & 4 & 8 \\
0 & 8 & 4 \\
\end{pmatrix}
,
\end{align}
when the Higgs field corresponds to $A^1$,
\begin{align}
\kappa_{ab2}=
\begin{pmatrix}
0 & 4 & 8 \\
4 & 2 & 6 \\
8 & 6 & 6 \\
\end{pmatrix}
,
\end{align}
when the Higgs field corresponds to $A^2$, and 
\begin{align}
\kappa_{ab3}=
\begin{pmatrix}
0 & 8 & 4 \\
8 & 6 & 6 \\
4 & 6 & 2 \\
\end{pmatrix}
,
\end{align}
when the Higgs field corresponds to $A^3$.

\begin{figure}[H]
	\centering
	\includegraphics[width=0.7\linewidth]{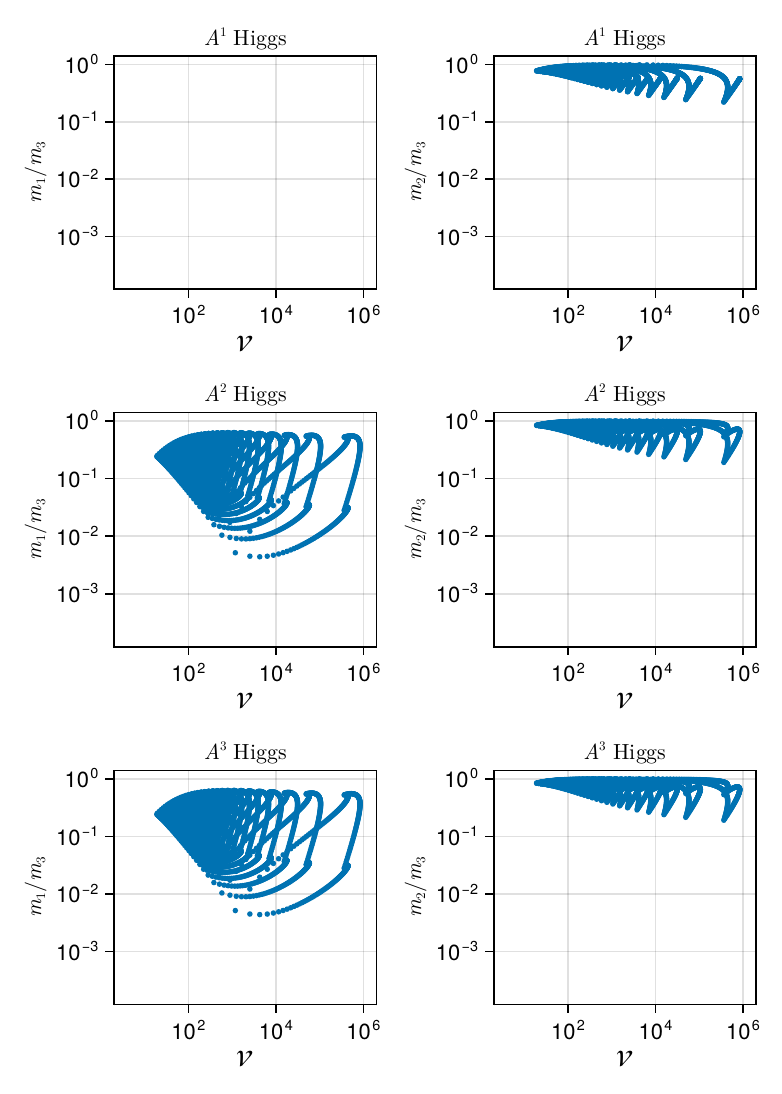}
	\caption{Mass ratios for CICY 6771 as a function of the minimum volume ${\cal V}$ where all real parts of the moduli are greater than or equal to 1. The masses are ordered as $m_1 < m_2 < m_3$. The left (right) panel shows the scatter for $m_1/m_3$ ($m_2/m_3$).}
	\label{fig:h11_3_moduli_3_V_to_ratio_type_3}
\end{figure}

Figure \ref{fig:h11_3_moduli_3_V_to_ratio_type_3} shows the mass ratios $m_1/m_3$ and $m_2/m_3$, which are obtained from eigenvalue ratios of $Y_{\hat a \hat b 1}$, $Y_{\hat a \hat b 2}$ and $Y_{\hat a \hat b 3}$. The points correspond to discrete variations of the moduli ratio $T^1:T^2:T^3$, subject to the restriction ${\rm Re}\,T^a\ge 1$.
In this example, the $(3\times 3)$ matrix $Y_{\hat a \hat b 1}$ has rank 1.
Thus, we have $m_1=0$ for any moduli values when the Higgs field corresponds to $A^1$.

\vspace{0.5cm}
$\bullet$~{\bf Type 4 (CICY 7069)}

As an illustrative example of Type 4, we examine CICY 7069.
The holomorphic Yukawa matrices are obtained from the intersection numbers as 
\begin{align}
\kappa_{ab1}=
\begin{pmatrix}
0 & 0 & 0 \\
0 & 2 & 6 \\
0 & 6 & 4 \\
\end{pmatrix}
,
\end{align}
when the Higgs field corresponds to $A^1$,
\begin{align}
\kappa_{ab2}=
\begin{pmatrix}
0 & 2 & 6 \\
2 & 0 & 2 \\
6 & 2 & 6 \\
\end{pmatrix}
,
\end{align}
when the Higgs field corresponds to $A^2$, and 
\begin{align}
\kappa_{ab3}=
\begin{pmatrix}
0 & 6 & 4 \\
6 & 2 & 6 \\
4 & 6 & 4 \\
\end{pmatrix}
,
\end{align}
when the Higgs field corresponds to $A^3$.

\begin{figure}[H]
	\centering
	\includegraphics[width=0.7\linewidth]{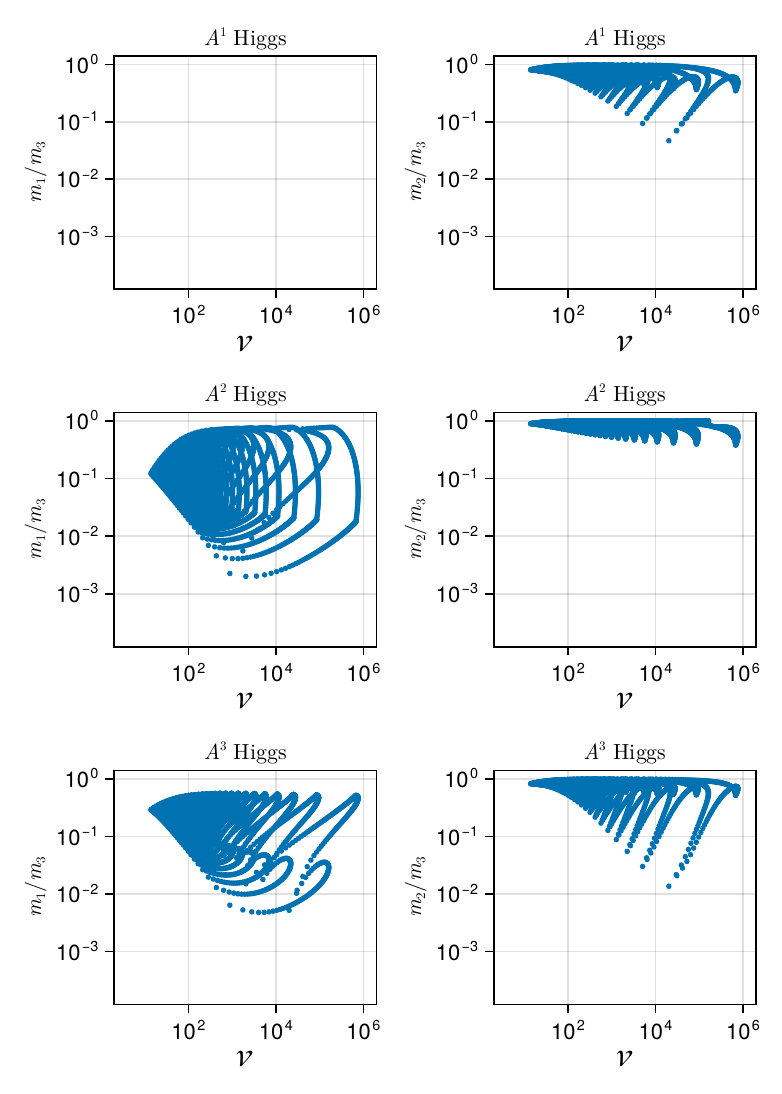}
	\caption{Mass ratios for CICY 7069 as a function of the minimum volume ${\cal V}$ where all real parts of the moduli are greater than or equal to 1. The masses are ordered as $m_1 < m_2 < m_3$. The left (right) panel shows the scatter for $m_1/m_3$ ($m_2/m_3$).}
	\label{fig:h11_3_moduli_3_V_to_ratio_type_4}
\end{figure}

Figure \ref{fig:h11_3_moduli_3_V_to_ratio_type_4} shows the mass ratios $m_1/m_3$ and $m_2/m_3$, which are obtained from eigenvalue ratios of $Y_{\hat a \hat b 1}$, $Y_{\hat a \hat b 2}$ and $Y_{\hat a \hat b 3}$. The points correspond to discrete variations of the moduli ratio $T^1:T^2:T^3$, subject to the restriction ${\rm Re}\,T^a\ge 1$.
In this example, the $(3\times 3)$ matrix $Y_{\hat a \hat b 1}$ has rank 1.
When the Higgs field corresponds to $A^1$, we have $m_1=0$ for any moduli values.

\vspace{0.5cm}
$\bullet$~{\bf Type 5 (CICY 7071)}

As an illustrative example of Type 5, we examine CICY 7071.
The holomorphic Yukawa matrices are obtained from the intersection numbers as 
\begin{align}
\kappa_{ab1}=
\begin{pmatrix}
0 & 2 & 4 \\
2 & 4 & 10 \\
4 & 10 & 8 \\
\end{pmatrix}
,
\end{align}
when the Higgs field corresponds to $A^1$,
\begin{align}
\kappa_{ab2}=
\begin{pmatrix}
2 & 4 & 10 \\
4 & 2 & 8 \\
10 & 8 & 10 \\
\end{pmatrix}
,
\end{align}
when the Higgs field corresponds to $A^2$, and 
\begin{align}
\kappa_{ab3}=
\begin{pmatrix}
4 & 10 & 8 \\
10 & 8 & 10 \\
8 & 10 & 4 \\
\end{pmatrix}
,
\end{align}
when the Higgs field corresponds to $A^3$.

\begin{figure}[H]
	\centering
	\includegraphics[width=0.7\linewidth]{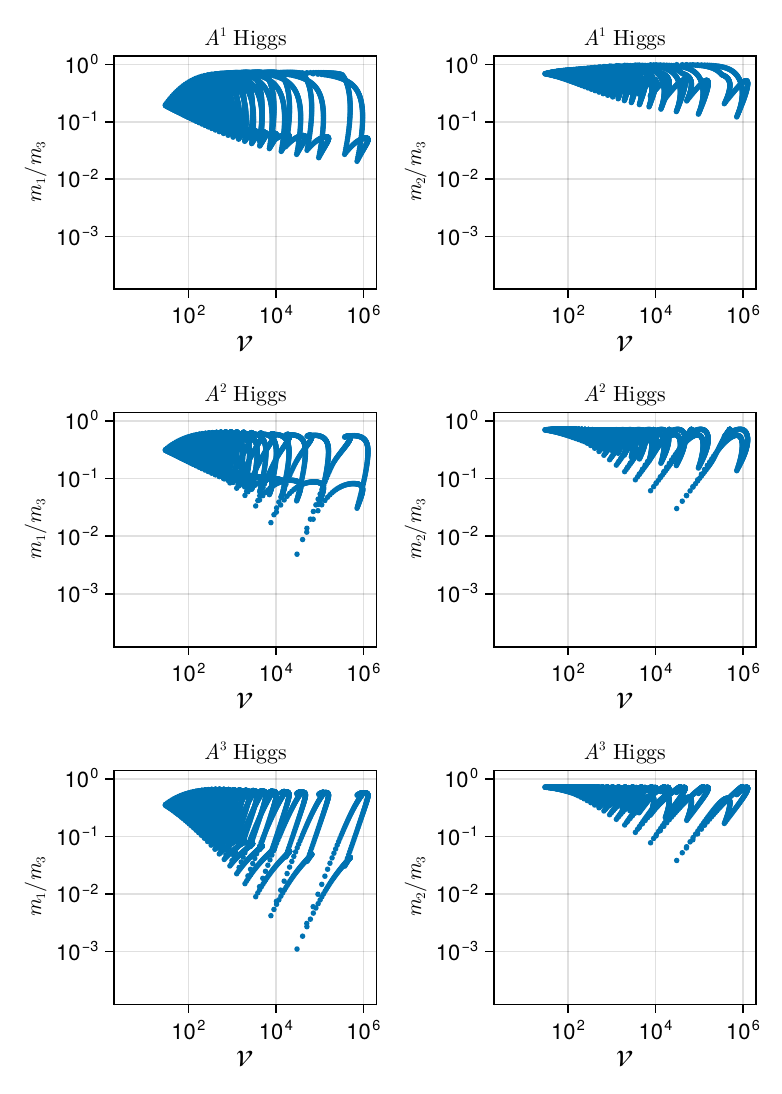}
	\caption{Mass ratios for CICY 7071 as a function of the minimum volume ${\cal V}$ where all real parts of the moduli are greater than or equal to 1. The masses are ordered as $m_1 < m_2 < m_3$. The left (right) panel shows the scatter for $m_1/m_3$ ($m_2/m_3$).}
	\label{fig:h11_3_moduli_3_V_to_ratio_type_5}
\end{figure}

Figure \ref{fig:h11_3_moduli_3_V_to_ratio_type_5} shows the mass ratios $m_1/m_3$ and $m_2/m_3$, which are obtained from eigenvalue ratios of $Y_{\hat a \hat b 1}$, $Y_{\hat a \hat b 2}$ and $Y_{\hat a \hat b 3}$. The points correspond to discrete variations of the moduli ratio $T^1:T^2:T^3$, subject to the restriction ${\rm Re}\,T^a\ge 1$.

\vspace{0.5cm}
$\bullet$~{\bf Type 7 (CICY 7450)}

As an illustrative example of Type 7, we examine CICY 7450.
The holomorphic Yukawa matrices are obtained from the intersection numbers as 
\begin{align}
\kappa_{ab1}=
\begin{pmatrix}
0 & 0 & 0 \\
0 & 0 & 4 \\
0 & 4 & 8 \\
\end{pmatrix}
,
\end{align}
when the Higgs field corresponds to $A^1$,
\begin{align}
\kappa_{ab2}=
\begin{pmatrix}
0 & 0 & 4 \\
0 & 0 & 0 \\
4 & 0 & 8 \\
\end{pmatrix}
,
\end{align}
when the Higgs field corresponds to $A^2$, and 
\begin{align}
\kappa_{ab3}=
\begin{pmatrix}
0 & 4 & 8 \\
4 & 0 & 8 \\
8 & 8 & 8 \\
\end{pmatrix}
,
\end{align}
when the Higgs field corresponds to $A^3$.

\begin{figure}[H]
	\centering
	\includegraphics[width=0.7\linewidth]{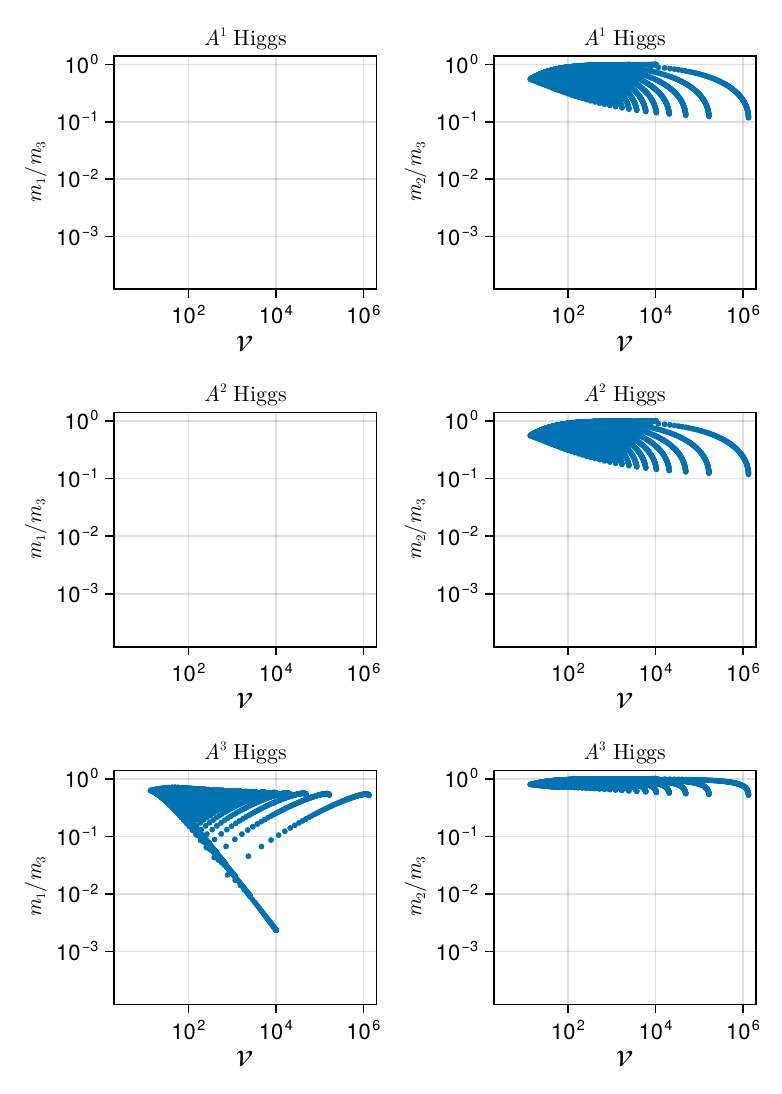}
	\caption{Mass ratios for CICY 7450 as a function of the minimum volume ${\cal V}$ where all real parts of the moduli are greater than or equal to 1. The masses are ordered as $m_1 < m_2 < m_3$. The left (right) panel shows the scatter for $m_1/m_3$ ($m_2/m_3$).}
	\label{fig:h11_3_moduli_3_V_to_ratio_type_7}
\end{figure}

Figure \ref{fig:h11_3_moduli_3_V_to_ratio_type_7} shows the mass ratios $m_1/m_3$ and $m_2/m_3$, which are obtained from eigenvalue ratios of $Y_{\hat a \hat b 1}$, $Y_{\hat a \hat b 2}$ and $Y_{\hat a \hat b 3}$. The points correspond to discrete variations of the moduli ratio $T^1:T^2:T^3$, subject to the restriction ${\rm Re}\,T^a\ge 1$.
In this example, the $(3\times 3)$ matrices  $Y_{\hat a \hat b 1}$ and $Y_{\hat a \hat b 2}$ have rank 1.
When the Higgs field corresponds to $A^1$ and $A^2$, we have $m_1=0$ for any moduli values.

\vspace{0.5cm}
$\bullet$~{\bf Type 8 (CICY 7465)}

As an illustrative example of Type 8, we examine CICY 7465.
The holomorphic Yukawa matrices are obtained from the intersection numbers as 
\begin{align}
\kappa_{ab1}=
\begin{pmatrix}
0 & 0 & 0 \\
0 & 0 & 4 \\
0 & 4 & 8 \\
\end{pmatrix}
,
\end{align}
when the Higgs field corresponds to $A^1$,
\begin{align}
\kappa_{ab2}=
\begin{pmatrix}
0 & 0 & 4 \\
0 & 0 & 4 \\
4 & 4 & 12 \\
\end{pmatrix}
,
\end{align}
when the Higgs field corresponds to $A^2$, and 
\begin{align}
\kappa_{ab3}=
\begin{pmatrix}
0 & 4 & 8 \\
4 & 4 & 12 \\
8 & 12 & 8 \\
\end{pmatrix}
,
\end{align}
when the Higgs field corresponds to $A^3$.

\begin{figure}[H]
	\centering
	\includegraphics[width=0.7\linewidth]{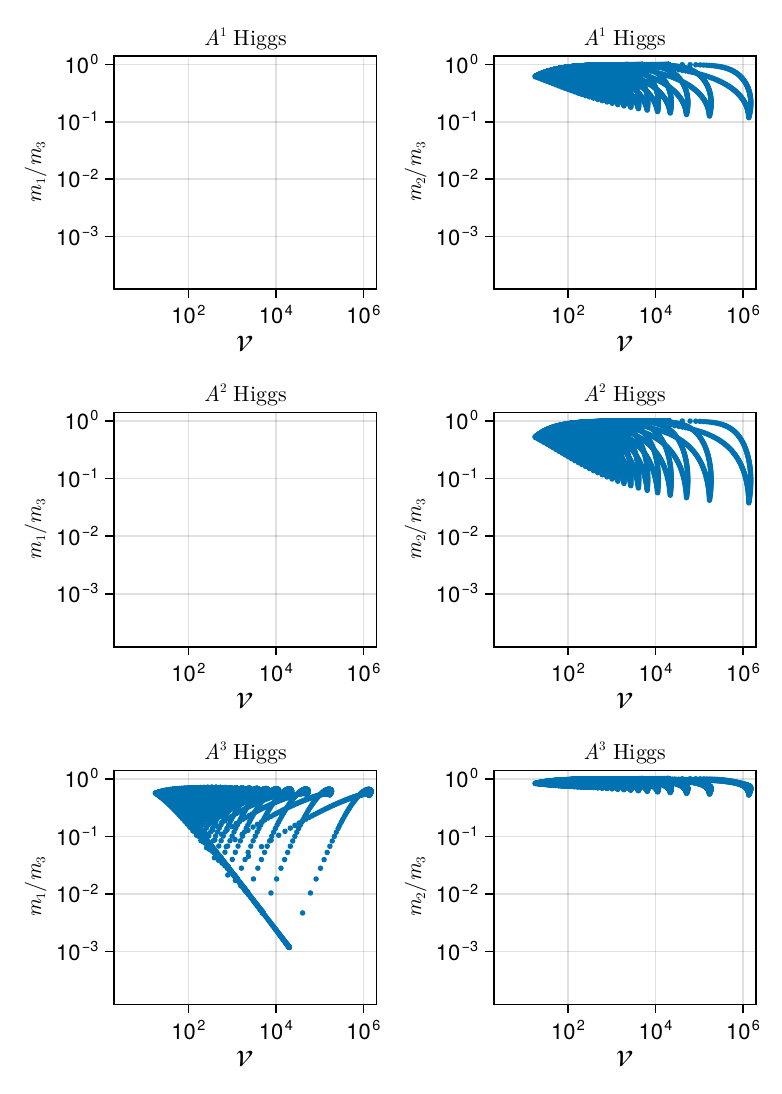}
	\caption{Mass ratios for CICY 7465 as a function of the minimum volume ${\cal V}$ where all real parts of the moduli are greater than or equal to 1. The masses are ordered as $m_1 < m_2 < m_3$. The left (right) panel shows the scatter for $m_1/m_3$ ($m_2/m_3$).}
	\label{fig:h11_3_moduli_3_V_to_ratio_type_8}
\end{figure}

Figure \ref{fig:h11_3_moduli_3_V_to_ratio_type_8} shows the mass ratios $m_1/m_3$ and $m_2/m_3$, which are obtained from eigenvalue ratios of $Y_{\hat a \hat b 1}$, $Y_{\hat a \hat b 2}$ and $Y_{\hat a \hat b 3}$. The points correspond to discrete variations of the moduli ratio $T^1:T^2:T^3$, subject to the restriction ${\rm Re}\,T^a\ge 1$.
In this example, the $(3\times 3)$ matrices  $Y_{\hat a \hat b 1}$ and $Y_{\hat a \hat b 2}$ have rank 1.
When the Higgs field corresponds to $A^1$ and $A^2$, we have $m_1=0$ for any moduli values.

\vspace{0.5cm}
$\bullet$~{\bf Type 9 (CICY 7708)}

As an illustrative example of Type 9, we examine CICY 7708.
The holomorphic Yukawa matrices are obtained from the intersection numbers as 
\begin{align}
\kappa_{ab1}=
\begin{pmatrix}
0 & 0 & 0 \\
0 & 2 & 5 \\
0 & 5 & 2 \\
\end{pmatrix}
,
\end{align}
when the Higgs field corresponds to $A^1$,
\begin{align}
\kappa_{ab2}=
\begin{pmatrix}
0 & 2 & 5 \\
2 & 0 & 3 \\
5 & 3 & 3 \\
\end{pmatrix}
,
\end{align}
when the Higgs field corresponds to $A^2$, and 
\begin{align}
\kappa_{ab3}=
\begin{pmatrix}
0 & 5 & 2 \\
5 & 3 & 3 \\
2 & 3 & 0 \\
\end{pmatrix}
,
\end{align}
when the Higgs field corresponds to $A^3$.

\begin{figure}[H]
	\centering
	\includegraphics[width=0.7\linewidth]{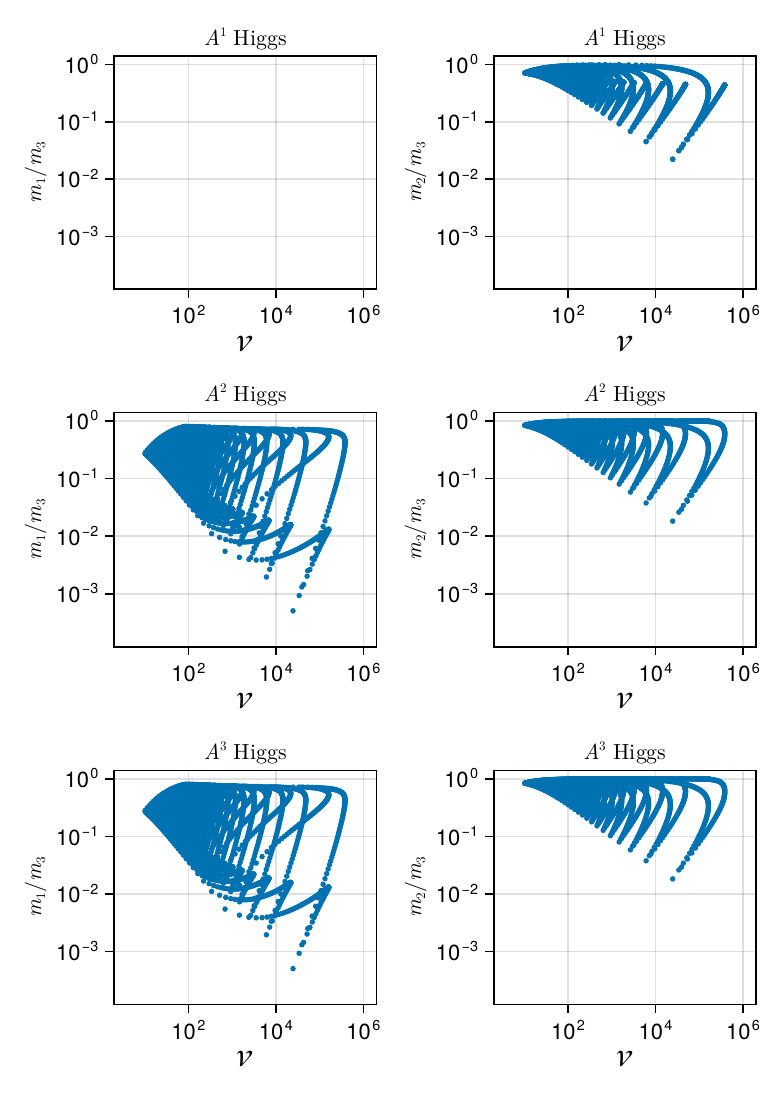}
	\caption{Mass ratios for CICY 7708 as a function of the minimum volume ${\cal V}$ where all real parts of the moduli are greater than or equal to 1. The masses are ordered as $m_1 < m_2 < m_3$. The left (right) panel shows the scatter for $m_1/m_3$ ($m_2/m_3$).}
	\label{fig:h11_3_moduli_3_V_to_ratio_type_9}
\end{figure}

Figure \ref{fig:h11_3_moduli_3_V_to_ratio_type_9} shows the mass ratios $m_1/m_3$ and $m_2/m_3$, which are obtained from eigenvalue ratios of $Y_{\hat a \hat b 1}$, $Y_{\hat a \hat b 2}$ and $Y_{\hat a \hat b 3}$. The points correspond to discrete variations of the moduli ratio $T^1:T^2:T^3$, subject to the restriction ${\rm Re}\,T^a\ge 1$.
In this example, the $(3\times 3)$ matrix $Y_{\hat a \hat b 1}$ has rank 1.
When the Higgs field corresponds to $A^1$, we have $m_1=0$ for any moduli values.

\vspace{0.5cm}
$\bullet$~{\bf Type 11 (CICY 7880)}

As an illustrative example of Type 11, we examine CICY 7880.
The holomorphic Yukawa matrices are obtained from the intersection numbers as 
\begin{align}
\kappa_{ab1}=
\begin{pmatrix}
0 & 0 & 0 \\
0 & 0 & 3 \\
0 & 3 & 2 \\
\end{pmatrix}
,
\end{align}
when the Higgs field corresponds to $A^1$,
\begin{align}
\kappa_{ab2}=
\begin{pmatrix}
0 & 0 & 3 \\
0 & 0 & 0 \\
3 & 0 & 2 \\
\end{pmatrix}
,
\end{align}
when the Higgs field corresponds to $A^2$, and 
\begin{align}
\kappa_{ab3}=
\begin{pmatrix}
0 & 3 & 2 \\
3 & 0 & 2 \\
2 & 2 & 0 \\
\end{pmatrix}
,
\end{align}
when the Higgs field corresponds to $A^3$.

\begin{figure}[H]
	\centering
	\includegraphics[width=0.7\linewidth]{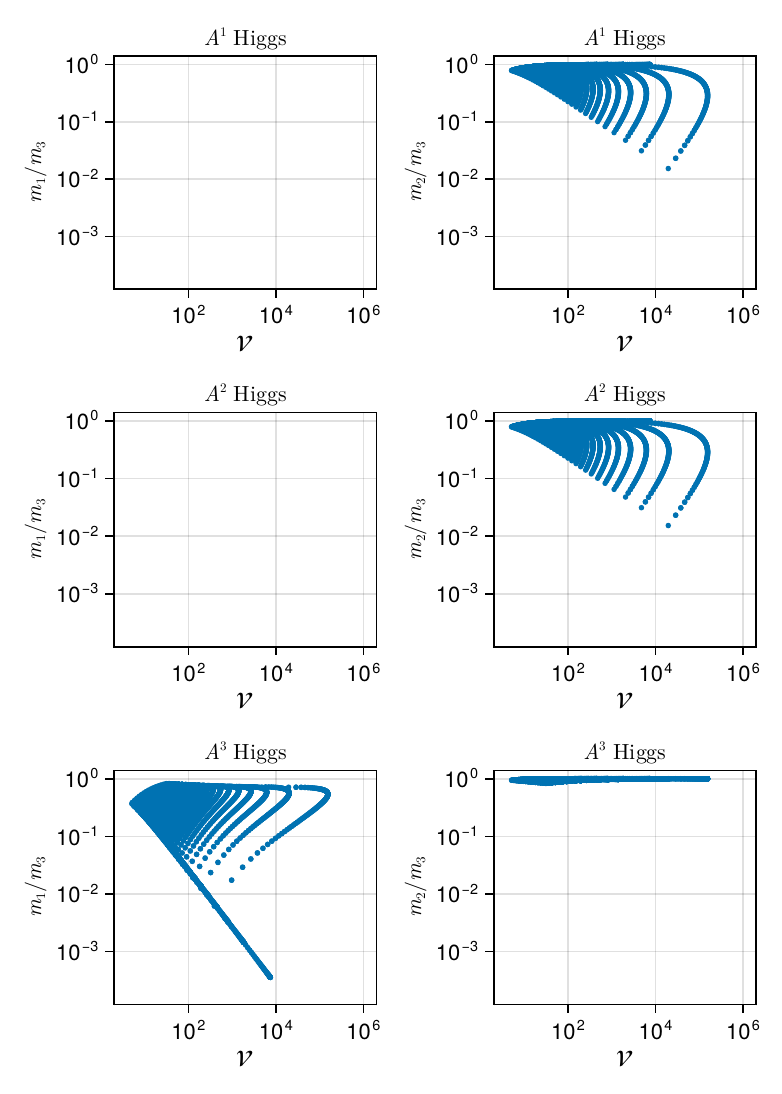}
	\caption{Mass ratios for CICY 7880 as a function of the minimum volume ${\cal V}$ where all real parts of the moduli are greater than or equal to 1. The masses are ordered as $m_1 < m_2 < m_3$. The left (right) panel shows the scatter for $m_1/m_3$ ($m_2/m_3$).}
	\label{fig:h11_3_moduli_3_V_to_ratio_type_11}
\end{figure}

Figure \ref{fig:h11_3_moduli_3_V_to_ratio_type_11} shows the mass ratios $m_1/m_3$ and $m_2/m_3$, which are obtained from eigenvalue ratios of $Y_{\hat a \hat b 1}$, $Y_{\hat a \hat b 2}$ and $Y_{\hat a \hat b 3}$. The points correspond to discrete variations of the moduli ratio $T^1:T^2:T^3$, subject to the restriction ${\rm Re}\,T^a\ge 1$.
In this example, the $(3\times 3)$ matrices  $Y_{\hat a \hat b 1}$ and $Y_{\hat a \hat b 2}$ have rank 1.
When the Higgs field corresponds to $A^1$ and $A^2$, we have $m_1=0$ for any moduli values.

\bibliography{references}{}
\bibliographystyle{JHEP}

\end{document}